\documentclass[11pt,a4paper]{article}

\providecommand{\newblock}{\hskip .11em plus.33em minus.07em}

\usepackage[margin=25mm]{geometry}
\usepackage{authblk}
\usepackage[numbers,sort&compress]{natbib}
\usepackage{booktabs}
\usepackage{enumitem}

\ifdefined\pdfminorversion\pdfminorversion=7\fi

\usepackage{physics}
\usepackage{amsmath,amssymb,amsthm}
\usepackage{mathtools}
\usepackage{graphicx}
\usepackage{xcolor}
\usepackage{quantikz}

\usepackage{subcaption}
\usepackage{algpseudocode}
\usepackage{algorithm}

\usepackage[hidelinks]{hyperref}
\usepackage{cleveref}
\crefname{subfigure}{fig.}{figs.}
\Crefname{subfigure}{Figure}{Figures}

\begin{document}
\title{Defect-Aware Parallel Atom Reloading Protocol for Neutral-Atom Quantum Computers~\thanks{A preliminary version of the defect-aware planner in this work appeared in Japanese in the Proceedings of the 40th Annual Conference of the Japanese Society for Artificial Intelligence~\cite{cite:Aoyama_JSAI2026}.
    This paper substantially extends the preliminary version by introducing a coherent parallel atom-reloading operation, reformulating and theoretically analyzing the planning problem, and expanding the numerical evaluation.}
}
\author[1,2]{Koki Aoyama\thanks{Corresponding author: \href{mailto:kotamanegi84@gmail.com}{kotamanegi84@gmail.com}.}}
\author[1]{Fumihiko Ino}
\affil[1]{Graduate School of Information Science and Technology, \newline The University of Osaka, 1-5 Yamadaoka, Suita, Osaka 565-0871, Japan}
\affil[2]{ALGO ARTIS Corporation, WeWork Hanzomon PREX North 10F, 2-3-2 Kojimachi, Chiyoda-ku, Tokyo 102-0083, Japan}
\date{}
\maketitle
\begin{abstract}
In this work, we propose a defect-aware parallel atom reloading protocol that reloads atoms based on defect information to increase the atom filling rate.
  The proposed protocol is useful for neutral-atom quantum computers, where reloading atoms into the defects, or atom-loss positions, is mandatory for continuous operation against atom loss.
  Our protocol takes advantage of both parallelism and defect awareness to improve atom-reloading efficiency.
  The proposed protocol consists of the coherent atom reloading operation and the defect-aware planner.
  The former replenishes the defects with atoms preserving the quantum information coherently.
  This coherence allows us to adaptively choose the reloading strategy based on the current defect configuration.
  The latter determines the atom reloading pattern to efficiently replenish the defects.
  In numerical experiments assuming a low-noise environment, we confirmed that the average atom filling rate improved from $98.61\%$ to $99.94\%$ for a $36 \times 90$ atom array, matching the filling rate obtained with a mixed integer programming solver to within $0.001$ percentage points while planning each reload in under $0.1$ ms.
  These results demonstrate that defect-aware parallel reloading can achieve and sustain high atom filling rates while the performance of the defect-aware planner meets the real-time requirements of neutral-atom quantum computers.
\end{abstract}

\section{Introduction}

The neutral-atom system is one of the most promising platforms for quantum computing, mainly because of its scalability in qubit count~\cite{cite:6100_atoms, cite:3000_coherent} and the versatility in qubit topology~\cite{cite:logical_qc, cite:self_citation}.
Each atom serves as a qubit in the neutral-atom-based quantum computer.
The atoms are arranged in a two-dimensional lattice to enable unified control.
The unified control allows the parallel execution of quantum gates, which is a key feature of the neutral-atom system.
The atoms can be transported to arbitrary positions in the atom array using optical tweezers~\cite{cite:chew_tweezers2024}.
This transport capability enables the implementation of various qubit topologies.

Current studies~\cite{cite:RSA_estimation, cite:Chemistry_estimation} on resource estimation suggest that more than a day of continuous system operation is required for utility-scale computation.
For instance, the decryption of RSA-2048 cryptography is estimated to require 5 days of continuous operation on a superconducting quantum computer~\cite{cite:RSA_estimation}.
For a chemistry-related use case, the quantum simulation of FeMoco is estimated to require 4 days of continuous operation on a superconducting quantum computer~\cite{cite:Chemistry_estimation}.
While these estimates are based on superconducting quantum computers, the neutral-atom system is estimated to require a similar amount of continuous operation time for utility-scale computation~\cite{cite:Lukin_resource_analysis}.

The atom reloading procedure is mandatory for continuous system operation in the neutral-atom-based quantum computer.
During operation, some of the atoms are lost due to idle noise.
Although quantum error correction codes~\cite{cite:erasure_error_code, cite:loss_basics_code} allow some atoms to be temporarily lost, the data becomes corrupted when there are too many defects, or empty sites, in the atom array.
To avoid data loss, we need to replenish the defects with atoms when atoms are lost.
The reloading rate of the atoms must be high enough to compensate for the atom loss in order to sustain atom availability in the atom array.

To improve the reloading efficiency, several hardware demonstrations have been reported.
Li et al.~\cite{cite:Yb_atom_replacement} experimentally demonstrated fast, continuous loading of fresh atoms.
Chiu et al.~\cite{cite:3000_coherent} experimentally realized parallel reloading with a predetermined reloading pattern.
The efficiency can be further improved by utilizing defect information, instead of relying on a predetermined pattern, to guide the reloading process.
Several groups~\cite{cite:cycle_time_neutral_atom1, cite:neutral_atom_coherent_reuse} have experimentally demonstrated reloading methods that use defect information to select the sites to replenish.
These methods operate atom-by-atom or row-by-row, and thus their execution time grows at least linearly with the array side length.
These limitations motivate the development of a parallel defect-aware atom reloading protocol that takes advantage of both parallelism and defect awareness for efficient reloading performance.

In this work, we propose an efficient defect-aware atom-reloading method which replenishes the defects with atoms while preserving the quantum information coherently.
The proposed method integrates the coherent parallel atom-reloading operation and the defect-aware planner in a manner that maximizes the reloading efficiency.
The proposed operation replenishes all defects in target positions specified by the operation parameters.
The proposed operation employs logical SWAP gates, similar to SWAP-LDU~\cite{cite:SWAP-LDU}, to execute both replacing existing atoms with new atoms and replenishing empty sites in the atom array simultaneously.
The coherence of the operation ensures that the operation parameters can be chosen based on the current defect configuration without risking data loss on existing atoms.
To maximize the efficiency of the operation, we solve an optimization problem which models the atom reloading process.
We define an objective function used for parameter optimization as the number of defects that the operation can replenish.
For a given objective function, the proposed defect-aware planner constructs a solution with a greedy algorithm and improves the greedy solution with hill climbing within the real-time planning budget.
The greedy algorithm constructs a solution by incrementally selecting rows and columns with the largest number of covered defects.
The hill climbing algorithm~\cite{cite:local_search} verifies the local optimality of the greedy solution and improves the greedy solution if necessary.
This method enables us to find a feasible operation that efficiently fills defects, thereby improving the filling rate of the atom array.

Our numerical evaluation shows that the proposed method improves the average atom filling rate from $98.61\%$ to $99.94\%$ for a $36 \times 90$ storage atom array under low-noise conditions, and from $93.22\%$ to $97.18\%$ under high-noise conditions.
We evaluated the performance of the proposed defect-aware planner by comparing it with the approach using mixed integer programming (MIP), which finds the optimal operation parameter.
Compared with the MIP method, the proposed method reaches the same filling rate to within $0.017$ percentage points in the low-noise case and within $0.11$ percentage points in the high-noise case over the completed MIP settings, while planning each reload in less than $0.37$ ms over the full evaluated range.
The runtime of the MIP method, in contrast, grows rapidly with the atom array size.
Individual planning calls already exceed the $40$ ms budget for a $12 \times 30$ storage atom array with the SCIP~\cite{cite:SCIP} solver under high-noise conditions, and the evaluation times out for a $36 \times 90$ storage atom array.
Our greedy algorithm alone attains a nearly optimal solution, and the hill-climbing refinement supplies a small additional gain that grows with the array size.
These results show that the proposed method is effective in improving the atom reloading efficiency while maintaining the real-time performance required for controlling the neutral-atom system.
By combining coherent atom replacement with real-time defect-aware planning, the proposed protocol offers a path toward sustaining high qubit availability in continuously operated neutral-atom quantum computers.

Our contributions are as follows.
\begin{itemize}
  \item We propose a defect-aware parallel atom reloading protocol, which combines the proposed atom-reloading operation with proposed real-time defect-aware planning.
        The integration of the operation and the defect-aware planner enables the efficient atom-reloading to improve the atom filling rate of the storage atom array.
  \item We propose a coherent parallel atom-reloading operation that replenishes all defects within the target positions specified by the operation parameters.
        The coherence of the operation enables the arbitrary selection of the target positions for atom reloading since the data in the existing atoms is preserved.
  \item We propose a defect-aware planner that constructs a nearly optimal solution based on a greedy algorithm.
        We also provide a hill climbing algorithm that improves the solution to the local optimum if necessary.
  \item Our numerical experiments show that the proposed method improves the average atom filling rate compared to the conventional method~\cite{cite:3000_coherent}.
        The proposed method achieves both high atom filling rates and real-time planning performance.
\end{itemize}

\section{Preliminaries}

\begin{figure*}[t]
  \centering
  \includegraphics[keepaspectratio, width=0.9\linewidth]{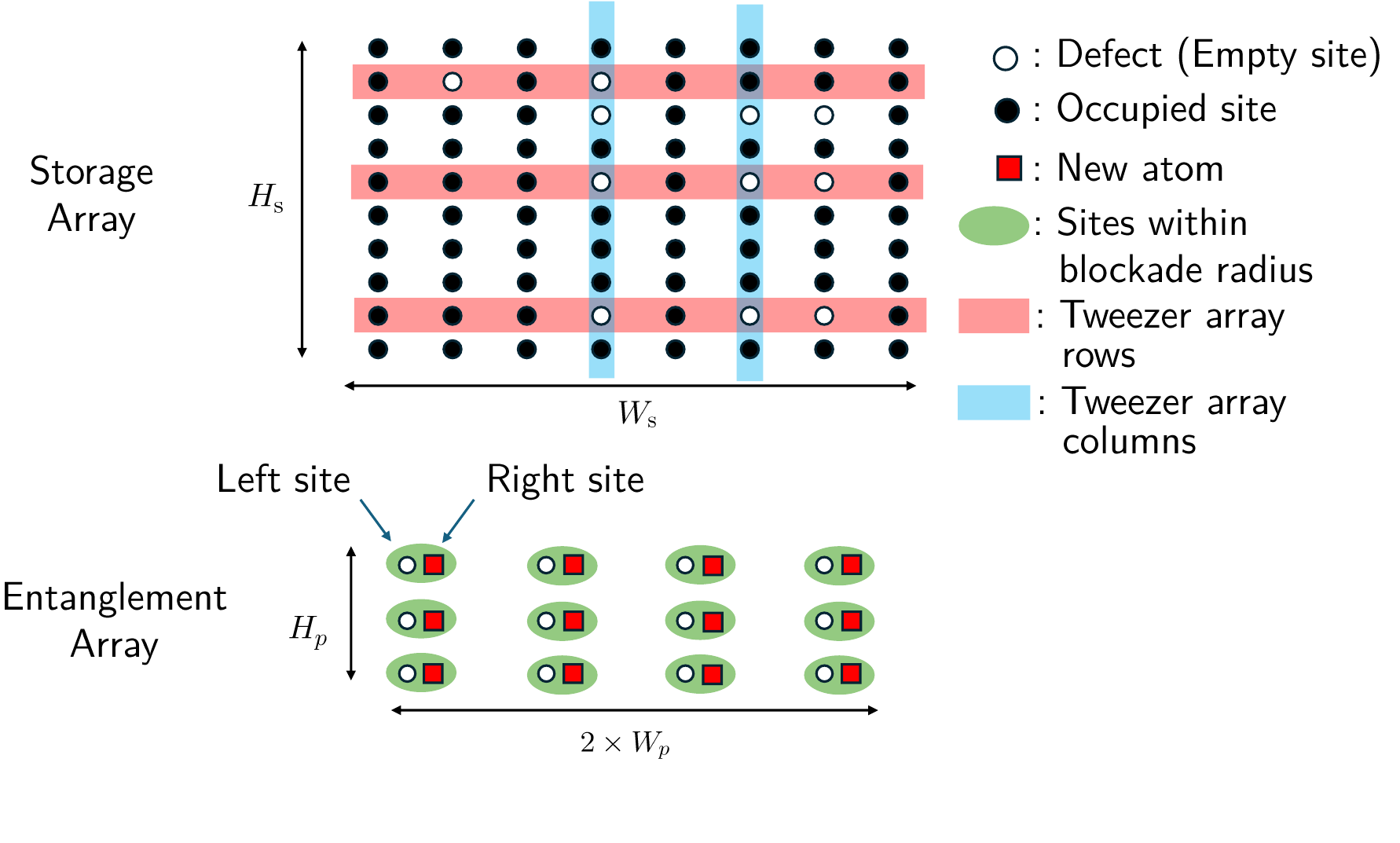}
  \caption{Schematic of the system model considered in this work.
    There are three components in the system: the storage atom array, the entanglement atom array, and the optical tweezer array.
    The storage array has $H_s \times W_s$ sites and holds atoms that store the intermediate computation data.
    Filled circles are sites occupied by atoms and open circles are defects, i.e., empty sites created by atom loss.
    The entanglement array consists of $H_p \times 2W_p$ sites grouped into $H_p \times W_p$ pairs.
    The two sites of a pair, referred to as the left site and the right site, are placed within the Rydberg blockade radius~\cite{cite:rydberg_review}, whereas sites belonging to different pairs are placed outside the blockade radius.
    This geometry allows the parallel execution of the CZ and H gates on all atoms in the entanglement array without crosstalk between different pairs.
    Before a reloading operation, newly prepared atoms occupy only the right site of each pair in the entanglement array.
    The optical tweezer array, generated by a pair of crossed AODs, transports atoms between the storage array and the entanglement array.
    The tweezers are generated at the intersection of a row set and a column set because the two AODs define a set of rows and a set of columns to be addressed, respectively.
    The figure illustrates one such choice of rows and columns by red and blue bands.
    Note that the addressed set contains occupied sites as well as defects, which is why the reloading operation should preserve the data held by those atoms.
  }
  \label{fig:neutral_atom_reloading_system}
\end{figure*}

In this work, we study the optical system, shown in \Cref{fig:neutral_atom_reloading_system}, with the following components.
\begin{itemize}
  \item \textbf{Storage atom array}: This storage atom array, or the storage array for short, stores atoms in use. The main focus of this work is to increase the number of available atoms throughout operation.
  \item \textbf{Entanglement atom array}: This entanglement atom array, or the entanglement array for short, stores temporary atoms to be reloaded into the storage array.
        The entanglement array is used to transport atoms to the storage array for reloading.
  \item \textbf{Optical tweezer array}: This optical tweezer array, or the tweezer array for short, transports atoms between the storage array and the entanglement array.
        The optical tweezer array is generated by a pair of crossed acousto-optic deflectors (AODs), which define a set of rows and a set of columns to be addressed, respectively.
\end{itemize}
We assume that the shape of every atom array is a two-dimensional rectangular lattice.
Let $H_s$ and $W_s$ denote the number of rows and columns of the storage array, respectively.
Similarly, let $H_p$ and $W_p$ denote the number of rows and columns of site pairs in the entanglement array, respectively.
The entanglement array has $H_p \times W_p$ pairs of atom sites, for a total of $H_p \times 2 W_p$ sites.

We intentionally design the entanglement array so that two atom sites forming a pair for a controlled-Z (CZ) gate are placed within the Rydberg blockade radius~\cite{cite:rydberg_review}, while all other atom sites are placed outside the blockade radius.
We refer to the two atom sites of each pair as the left site and the right site.
This geometry allows every pair to be simultaneously addressed for the CZ gate without crosstalk between different pairs.
With this design, parallel execution of the CZ and Hadamard (H) gates on all atoms in the entanglement array is enabled.

The goal of this work is to maximize the storage array's filling rate.
Atom loss remains permanent until the atoms are replenished from elsewhere.
For that reason, the operations that access the defects result in errors.
Hence, increasing the filling rate of the storage array leads to lower error rates in general.

The system repeats the following procedures to achieve the goal.
\begin{enumerate}
  \item \textbf{Atom loss detection procedure}: The system identifies the positions where atom loss occurred.
        In hardware, atom loss is detected based on the output of the measurement operation.
        The atom loss of a site can be directly observed when executing the measurement operation on that site.
        In quantum error correction, the atom loss can be inferred from the syndrome measurements~\cite{cite:Suzuki_stabilizer,cite:AI_assisted_decoding,cite:eraser,cite:transmon_leakage}.
        For simplicity, we assume that the atom loss is detected immediately after the loss occurs.
  \item \textbf{Atom reloading procedure}: Atoms are transferred from the entanglement array to replenish the defects in the storage array.
        The quantum information stored in the existing atoms should be preserved during this procedure.
\end{enumerate}

\subsection{Mathematical Representation of the System}

We represent the state of the storage atom array as a binary matrix $A$.
The matrix element $A_{i,j}$ in row $i$ and column $j$ is $1$ if and only if the lattice point at row $i$ and column $j$ holds an atom.
For instance, the following $2 \times 3$ binary matrix
\begin{equation}
  A = \begin{pmatrix}
    1 & 1 & 0 \\
    0 & 1 & 1 \\
  \end{pmatrix},
\end{equation}
represents a state which has atoms at the lattice points $(i,j) = (1,1), (1,2), (2,2)$, and $(2,3)$.

The system repeatedly reloads atoms while in operation.
We define $1$ unit time as the execution time of one atom reloading procedure.
We represent the state of the storage array at time $t$ as $A^{(t)}$.
The state is updated recursively from $A^{(t)}$ to $A^{(t+1)}$ with certain rules, which are described below.

The atom loss detection procedure applies the effect of idle noise to the storage array $A^{(t)}$.
In this work, we assume that the idle noise causes each atom in $A^{(t)}$ to disappear with identical probability $p_\mathrm{idle}$.
Formally, $A^{(t)}$ is updated to the new state $A'^{(t)}$ defined by
\begin{equation}
  A'^{(t)}_{i,j} = \begin{cases}
    0             & \text{(with probability $p_\mathrm{idle}$)}, \\
    A^{(t)}_{i,j} & \text{(otherwise)}.
  \end{cases}
\end{equation}

The atom reloading procedure replenishes the storage array $A'^{(t)}$ with atoms, producing $A^{(t+1)}$.
There are various ways to implement the atom reloading procedure.
The planner first determines which atoms to reload, and then the hardware executes the atom reloading procedure.
We assume that the time budget of the planner is $40$\,ms for each atom reloading procedure because the preparation and reconfiguration of the defect-free atom array, which can be executed in parallel with the planning, takes approximately $40$\,ms~\cite{cite:3000_coherent}.
Exceeding this time budget may result in the atom reloading procedure being skipped.
We also assume that the defect information is available at the start of a planner invocation.
The overall atom reloading procedure is executed every $80$\,ms~\cite{cite:3000_coherent} to maintain the filling rate of the storage array.

\subsection{Objective of the Atom Reloading Procedure}
The objective of the atom reloading procedure is to maximize the filling rate $F$ of $A$, specified by
\begin{equation}\label{eq:filling_rate}
  F = \frac{1}{(T-C) \cdot H_s \cdot W_s} \sum^T_{t=C+1} \sum^{H_s}_{i = 1} \sum^{W_s}_{j = 1} A^{(t)}_{i,j},
\end{equation}
for a total number of time steps $T$ and the burn-in period $C$.
To mitigate the effect of the initial state, we discard the first $C$ time steps as a burn-in period and compute the filling rate over the remaining steps.

\subsection{Basic Operations}

We assume that the following basic operations, with parameters $I$ and $J$ representing the row and column indices of the storage array, are supported on hardware.
\begin{itemize}
  \item \textbf{Preparation}:
        Load atoms into the entanglement array.
        In this operation, atoms are loaded into the right site of each site pair in the entanglement array.
        The atoms are reset to the $\ket{0}$ state.
        This is experimentally implemented in~\cite{cite:3000_coherent,cite:atom_by_atom_assembler}.
  \item \textbf{Transport}: Transport the atoms between the entanglement array and the positions in the storage array specified by $I \times J$ using the optical tweezer array.
        This is experimentally implemented in~\cite{cite:3000_coherent}.
  \item \textbf{H-gate}: Apply the H gate to all existing atoms in the entanglement array.
        This operation changes the basis of the qubits between $\{\ket{0},\ket{1}\}$ and $\{\ket{+},\ket{-}\}$.
        This is experimentally implemented in~\cite{cite:3000_coherent,cite:CZ_parallel}.
  \item \textbf{CZ-gate}: Apply the CZ gate to all pairs of atoms in the entanglement array.
        An empty site is equivalent to a site occupied by an atom in the $\ket{0}$ state.
        In the standard parallel CZ protocol~\cite{cite:CZ_PRL}, the global Rydberg drive couples only $\ket{1}$ to the Rydberg state.
        Hence, a partner in $\ket{0}$ and an absent partner are indistinguishable from the viewpoint of the other atom in the pair.
        In both cases, the other atom in the pair follows the unblockaded single-atom trajectory and returns to its qubit state with the single-atom light-shift phase, which is common to all atoms and is compensated as part of the standard protocol.
        The CZ gate therefore acts as the identity on the remaining atom of a pair whose partner is missing.
  \item \textbf{Discard}: Discard the atoms in the entanglement array.
\end{itemize}

\subsection{Leakage Detection Unit}

The leakage detection unit (LDU) identifies atoms that have leaked out of the computational subspace or been lost from the array.
The LDU typically operates by applying a sequence of quantum gates and measurements that map leakage or loss information onto ancilla qubits.
The information obtained by the LDU allows us to take corrective actions, such as reloading lost atoms or applying error-correction protocols to mitigate the effects of leakage.

The SWAP-LDU~\cite{cite:SWAP-LDU} is a specific implementation of the LDU that employs a SWAP operation to detect and identify leakage coherently.
The SWAP-LDU first transfers the state of the target qubit to an ancilla qubit using a SWAP operation.
The ancilla qubit becomes the new owner of the quantum information, while the original target qubit can be measured to detect leakage without disturbing the logical state.
This approach allows for coherent leakage detection because the logical information is preserved in the ancilla qubit while the target qubit is probed for leakage.

The advantage of the SWAP-LDU is that it reloads lost atoms while simultaneously detecting leakage.
Unlike other LDU implementations, the SWAP-LDU replaces leaked or lost atoms with fresh ones.
Extending the concept of the SWAP-LDU to multiple qubits allows us to detect leakage and reload lost atoms across the array.

Baranes et al.~\cite{cite:SWAP-SE} investigated the integration of loss detection and qubit replacement into syndrome extraction.
In the quantum error correction context, the syndrome extraction can be used to detect errors including qubit loss.
The SWAP-SE scheme~\cite{cite:SWAP-SE} allows for the simultaneous detection of qubit loss and the replacement of lost qubits during syndrome extraction.
This integration allows for optimized handling of qubit loss through circuit-level optimization.
Compared to this approach, our work addresses coherent parallel atom reloading and the optimization of reloading patterns under the geometric constraints of AOD-based transport.

\subsection{Challenge in the Parallel Atom Reloading Procedure}
\label{sec:geometric_constraint}

The key challenge in the parallel atom reloading procedure is to find efficient reloading strategies while complying with the geometric constraint on the transport operation.
The geometric constraint originates from the way the mobile tweezer array is generated.
In common hardware settings~\cite{cite:3000_coherent,cite:logical_qc}, the storage array is held by static traps, while the mobile tweezers used for transport are generated by a pair of crossed AODs.
Each AOD is driven by a multi-tone radio-frequency signal in which each tone deflects the beam to one angle, that is, to one coordinate along the corresponding axis.
One AOD therefore defines a set of rows and the other defines a set of columns, and the resulting light field is the product of the two.
Consequently, the set of sites addressable in a single parallel transport is always the Cartesian product $I \times J$ of a row set $I$ and a column set $J$.

This product structure makes the atom-reloading problem non-trivial.
If an arbitrary subset of sites could be addressed at once, the optimal strategy would be simply to address the defect set itself, and no planning would be required.
However, addressing an arbitrary subset instead requires either a hologram-based trap array~\cite{cite:SLM_assist}, whose manipulation is slower than that of the AODs, or sequential atom-by-atom or row-by-row transport~\cite{cite:atom_by_atom_assembler,cite:cycle_time_neutral_atom1,cite:neutral_atom_coherent_reuse}, whose execution time grows at least linearly with the side length of the array.

The number of rows and columns that can be selected simultaneously is bounded by the number of radio-frequency tones each AOD can sustain.
In this work, we assume that the number of rows and columns is bounded by the entanglement array size $|I| \leq H_p$ and $|J| \leq W_p$.
The selected rows $I$ and columns $J$ may be arbitrary because the tone frequencies can be chosen independently.

\section{Related Work}

\subsection{Atom Reloading Procedure}

Chiu et al.~\cite{cite:3000_coherent} demonstrate a parallel atom reloading procedure that replenishes the storage array with atoms in a predetermined pattern.
The approach is to split the atoms into $6$ groups based on their row and column indices modulo $3$ and $2$, respectively.
Specifically, at time $t$, the procedure replaces atoms at sites in $I \times J$, where $I, J$ are set as
\begin{equation}
  \begin{aligned}
    I & = \{i \in \mathbb{Z} \mid 1 \leq i \leq H_s \land i \bmod 3 = t \bmod 3\},                  \\
    J & = \{j \in \mathbb{Z} \mid 1 \leq j \leq W_s \land j \bmod 2 = \lfloor t/3 \rfloor \bmod 2\}
  \end{aligned}
\end{equation}
The procedure first discards the existing atoms in the positions specified by $I$ and $J$.
Then, the algorithm prepares new atoms in the entanglement array.
Finally, the algorithm transfers the atoms in the entanglement array to the positions specified by $I$ and $J$ in the storage array.
The atom transfer is performed simultaneously in a single action using the optical tweezer array built on an acousto-optic deflector.
This allows us to replenish the storage array with atoms in parallel using a predetermined pattern.

The problem with this parallel atom reloading protocol is that defect information is not taken into account when determining the reloading pattern.
Even if all atoms in the specified positions are already present, the algorithm still discards them and reloads new atoms.
This behavior leads to inefficient atom reloading, resulting in a lower filling rate of the storage array.
The reloading performance can be improved if the algorithm can choose which rows and columns of the storage array to replenish based on defect information.

The difficulty of utilizing defect information in the parallel atom reloading procedure is that the operation discards all atoms, together with the data stored in atoms, in the specified rows and columns.
The data stored in the atoms is lost when they are discarded.
To avoid data loss, the algorithm must transfer the data in the affected atoms to other atoms before the operation.
If the rows and columns are predetermined well in advance before the operation, the algorithm can plan the data transfer in advance to avoid data loss.
However, if the rows and columns are determined based on real-time defect information, the algorithm cannot plan the data transfer in advance, which may result in data loss.
This motivates the development of a new parallel coherent atom reloading procedure that sustains the data stored in the discarded atoms while replenishing the defects.

Other studies~\cite{cite:cycle_time_neutral_atom1, cite:neutral_atom_coherent_reuse} experimentally demonstrated the defect-aware atom reloading.
In these studies, the defect-aware atom reloading is achieved by row-by-row or atom-by-atom replenishment.
We model their protocol as follows.
At time $t$, we set the row and column indices $(I,J)$ to
\begin{equation}\label{eq:row_by_row_indices}
  \begin{aligned}
    I & = \{i \in \mathbb{Z} \mid i = 1 + (t \bmod H_s) \},              \\
    J & = \{j \in \mathbb{Z} \mid A'^{(t)}_{1 + (t \bmod H_s),j} = 0 \},
  \end{aligned}
\end{equation}
so that all the defects in the row $(1 + (t \bmod H_s))$ are replenished.
In case $|J| > W_p$, we select a subset of $J$ with size $W_p$ to match the width of the entanglement array.
The algorithm transfers the atoms in the entanglement array to the positions specified by $I$ and $J$ in the storage array.

This protocol, however, does not scale well with the size of the atom array.
Because this atom reloading is performed atom-by-atom or row-by-row, execution time scales at least linearly with the side length of the atom array, limiting scalability.
This motivates the parallel defect-aware protocol proposed in this work.

\subsection{Atom Reconfiguration Procedure}

There are various methods~\cite{cite:atom_by_atom_assembler,cite:PSCA,cite:tetris,cite:SLM_assist,cite:self_citation} for atom reconfiguration, which rearranges stochastically loaded atoms to form a defect-free atom array.
A commonly used approach is the PSCA/Tetris algorithm~\cite{cite:PSCA,cite:tetris}, which reconfigures atoms via parallel row and column moves.
This algorithm first identifies the positions of the defects in the atom array.
Then, the algorithm performs row-by-row and column-by-column parallel moves to fill the defects in the atom array.
This strategy allows atoms in a single atom array to be reconfigured in $\mathrm{O}(H + W)$ operations where $H$ and $W$ are the height and width of the atom array, respectively.

The difference between the atom reloading procedure and the atom reconfiguration procedure is that the former replenishes the storage array with atoms, while the latter rearranges atoms in the atom array.
The atom reconfiguration procedure affects only the atom array.
This allows for the high-quality initialization of a defect-free atom array.
In contrast, the atom reloading procedure allows inter-atom-array operations, which affect both the storage array and the entanglement array.
This allows for the replenishment of the storage array with atoms, which contributes to the continuous operation of the system.
These differences highlight the distinct operational scopes of the two procedures.

\section{Proposed Method}

\begin{figure}
  \centering
  \includegraphics[keepaspectratio, width=0.98\linewidth]{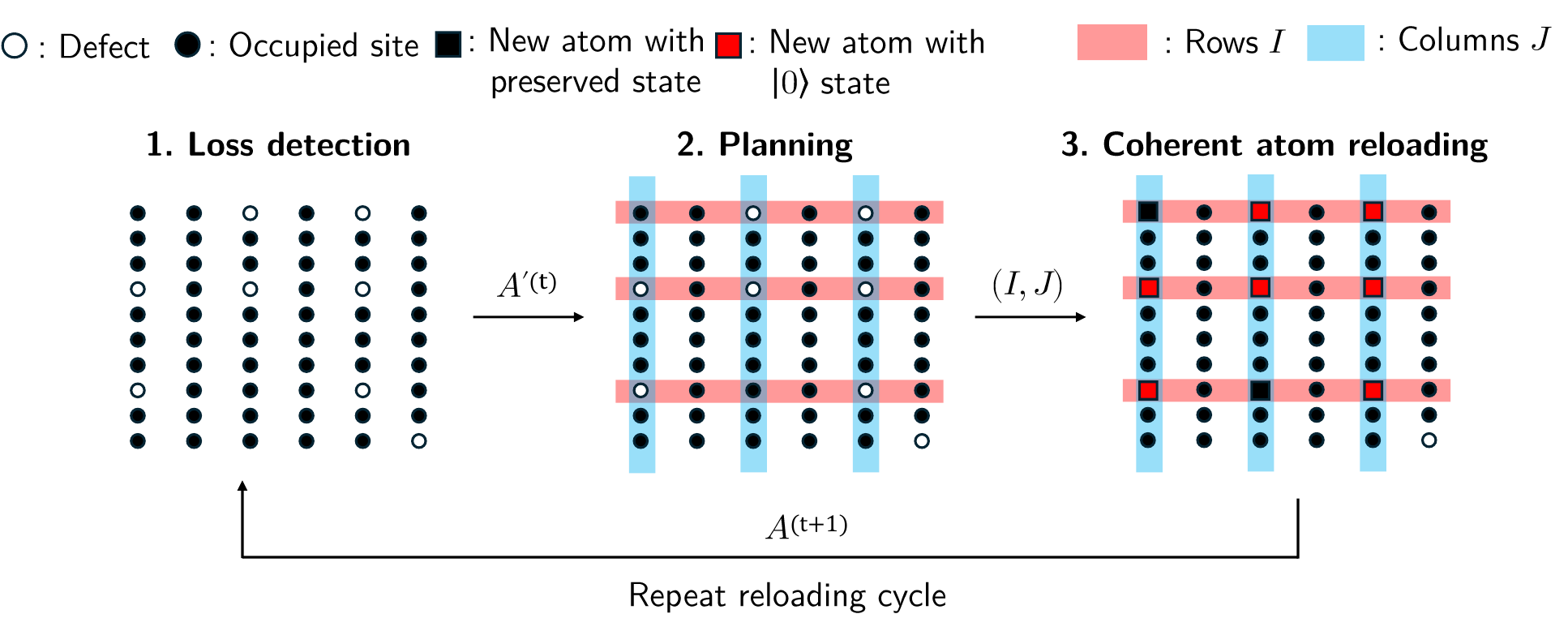}
  \caption{
  Overview of the proposed defect-aware atom reloading protocol.
  In step~1, the atom loss detection procedure locates the defects and yields the state $A'^{(t)}$.
  In step~2, the defect-aware planner determines the operation parameters $(I,J)$ based on the current state $A'^{(t)}$.
  The operation parameters $(I,J)$ are the row and column indices of the storage array to be replenished.
  Only the defects in the storage array specified by $I \times J$ are replenished with atoms because the tweezers are generated by AODs.
  In step~3, the coherent parallel atom reloading operation replenishes the defects in $I \times J$ with new atoms in the default state $\ket{0}$.
  The sites in $I \times J$ that already hold an atom are replaced with new atoms while preserving the quantum information stored in the existing atoms.
  The resulting state $A^{(t+1)}$ becomes the input of the next cycle.
  Selecting the operation parameters $(I,J)$ based on defect information allows the proposed method to improve the filling rate of the storage array.
  }
  \label{fig:proposed_method}
\end{figure}

The proposed method is composed of two components: the coherent parallel atom reloading operation and the defect-aware planner.
The overview of the proposed method is shown in \Cref{fig:proposed_method}.
The coherent parallel atom reloading operation is designed to replenish defects in the storage array with new atoms while preserving the data stored in the existing atoms.
Formally, the operation takes two parameters $(I,J)$, which are the row and column indices of the storage array to be replenished.
With the operation, the storage array is updated from $A'^{(t)}$ to $A^{(t+1)}$ using the following equation.
\begin{equation}\label{eq:state_update}
  A^{(t+1)}_{i,j} = \begin{cases}
    1              & \text{(if $i \in I$ and $j \in J$ and with probability $1-p_\mathrm{op}$)}, \\
    0              & \text{(if $i \in I$ and $j \in J$ and with probability $p_\mathrm{op}$)},   \\
    A'^{(t)}_{i,j} & \text{(if $i \notin I$ or $j \notin J$)}.
  \end{cases}
\end{equation}
The property of coherence allows the operation parameters to be determined based on defect information because the data in the existing atoms is preserved regardless of the chosen rows and columns $(I,J)$.
Our proposed defect-aware planner determines the operation parameters by solving the optimization problem whose objective is to maximize the number of defects that can be replenished in the storage array.
The proposed defect-aware planner first constructs a solution with a greedy algorithm and then improves the greedy solution with hill climbing.
The greedy algorithm has a time complexity of $\mathrm{O}(H_s W_s + (H_p + W_p)(H_s + W_s))$, which is sufficient to meet the real-time planning budget.
These designs allow the proposed method to adaptively choose which rows and columns of the storage array to replenish based on defect information, improving the filling rate of the storage array.

\subsection{Coherent Parallel Atom Reloading Operation}\label{sec:coherent_operation}

\begin{figure}[t]
  \centering
  \includegraphics[keepaspectratio, width=0.98\linewidth]{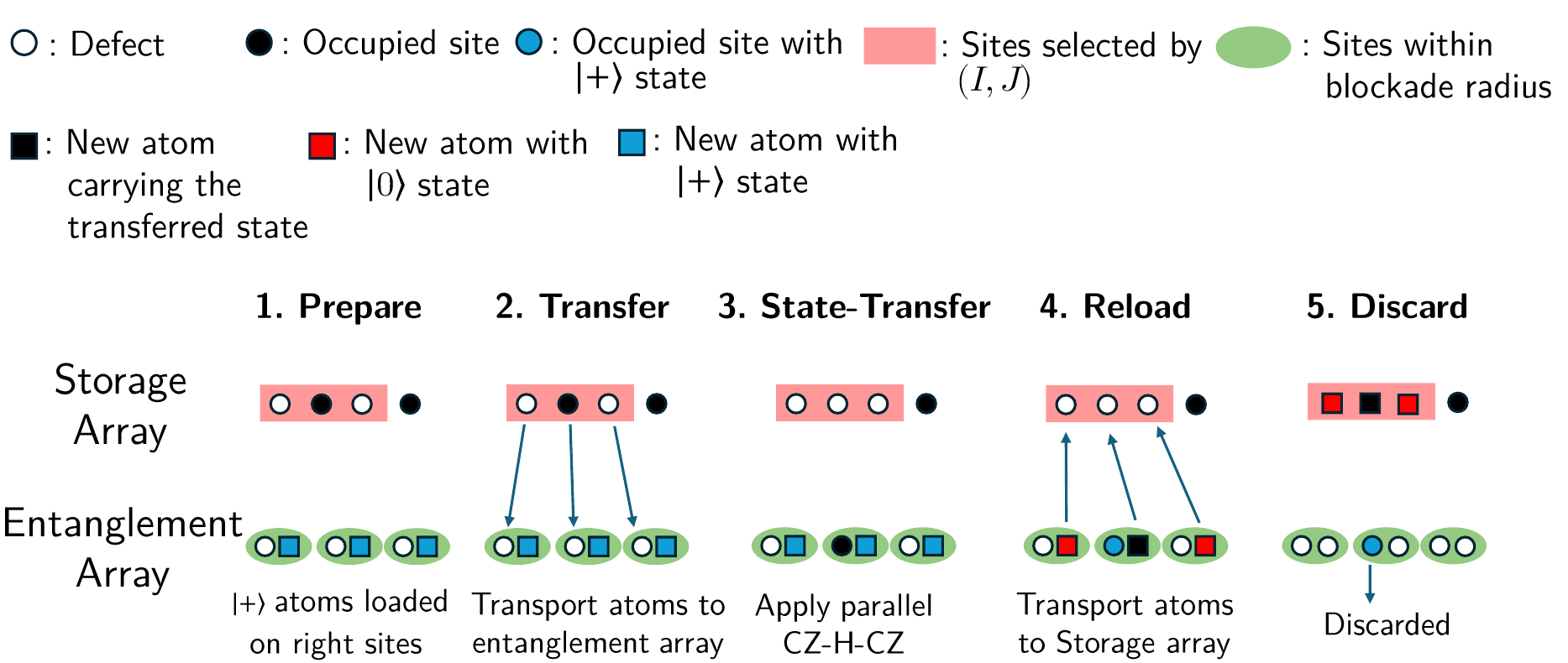}
  \caption{Overview of the coherent parallel atom-reloading operation.
    The operation takes two parameters $(I,J)$, which are the row and column indices of the storage array to be replenished.
    Each panel shows the state at the beginning of the corresponding operation, with the transport performed in that operation drawn as arrows.
    Step~1, the prepare operation, consists of the preparation operation followed by the H-gate operation and leaves a $\ket{+}$ atom on the right site of every pair.
    Steps~2 to~4 move the selected storage atoms into the left sites of the entanglement array, apply the parallel CZ-H-CZ sequence to transfer the data from the left sites to the right sites, and move the right-site atoms back into the storage array, respectively.
    Step~5, the discard operation, discards the remaining atoms in the left sites of the entanglement array.
    After these operations, the defects in the storage array are replenished with new atoms in the default state.
    The existing atoms are physically replaced with new atoms carrying the same data.
  }
  \label{fig:coherent_parallel_atom_reloading_operation}
\end{figure}

We propose a coherent parallel atom-reloading operation that operates on the selected rows and columns of the storage array.
The main idea of the coherent parallel atom-reloading operation is to extend the SWAP-LDU~\cite{cite:SWAP-LDU} to enable coherent and parallel reloading of atoms.
Because of the constraints on AODs, the coherent parallel atom-reloading operation must carefully coordinate the transport of atoms to avoid collisions.
The most critical aspect is to synchronize the movement of atoms in the storage array and the entanglement array to maintain coherence and parallelism.

We consider the atom-reloading operation depicted in \Cref{fig:coherent_parallel_atom_reloading_operation}.
The atom-reloading operation takes two parameters $(I,J)$, which are the row and column indices of the storage array to be replenished.
The following five steps implement the atom-reloading operation.
\begin{enumerate}
  \item \textbf{Prepare}: Prepare $\ket{+}$ state atoms in the right sites of the entanglement array using the preparation operation and the H-gate operation.
  \item \textbf{Transfer}: We execute the transport operation to physically transfer atoms in the storage array at the positions specified by $I$ and $J$ to the left sites of the entanglement array.
  \item \textbf{State-Transfer}: The parallel logical state-transfer operation transfers the data stored in the left sites of the entanglement array to the right sites of the entanglement array.
  \item \textbf{Reload}: Physically transfer the atoms in the right sites of the entanglement array to the corresponding positions in the storage array using the transport operation.
  \item \textbf{Discard}: Discard the remaining atoms in the entanglement array using the discard operation.
\end{enumerate}
After these steps, the new atoms with default data are transferred into the defects in the storage array, while existing atoms hold the same data.
Since this operation is executed in parallel without corrupting the data of the existing atoms, we call this operation a coherent parallel atom-reloading operation.

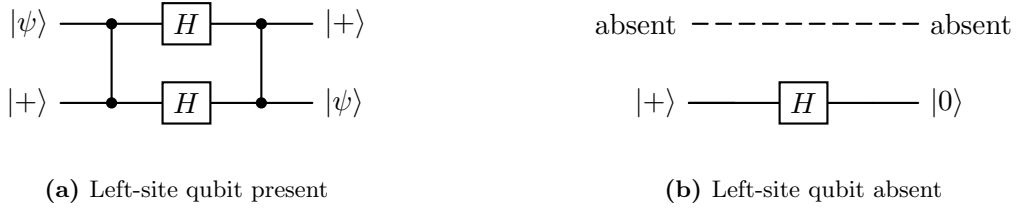
\begin{figure}[t]
  \centering
  \captionsetup[subfigure]{position=bottom, singlelinecheck=true}
  \begin{subfigure}[b]{0.49\linewidth}
    \centering
    \begin{quantikz}[row sep=0.5cm, column sep=0.6cm]
      \lstick{$\ket{\psi}$}
      & \ctrl{1} & \gate{H} & \ctrl{1}
      & \rstick{$\ket{+}$} \\
      \lstick{$\ket{+}$}
      & \control{} & \gate{H} & \control{}
      & \rstick{$\ket{\psi}$} \\
    \end{quantikz}
    \caption{Left-site qubit present}
    \label{fig:state_transfer_present}
  \end{subfigure}
  \hfill
  \begin{subfigure}[b]{0.49\linewidth}
    \centering
    \begin{quantikz}[row sep=0.5cm, column sep=0.6cm]
      \lstick{\textrm{absent}}
      & \ghost{H}\wireoverride{n} & \ghost{H}\wireoverride{n} & \ghost{H}\wireoverride{n}
      & \wireoverride{n}\wire[l][4][dashed]{q}\rstick{\textrm{absent}} \\
      \lstick{$\ket{+}$}
      & \qw & \gate{H} &
      & \rstick{$\ket{0}$} \\
    \end{quantikz}
    \caption{Left-site qubit absent}
    \label{fig:state_transfer_absent}
  \end{subfigure}
  \caption{Quantum circuit of the parallel logical state-transfer operation.
    \Cref{fig:state_transfer_present,fig:state_transfer_absent} show the cases where the left-site qubit is present and absent, respectively.
    In~\Cref{fig:state_transfer_present}, the gate sequence $\mathrm{CZ}\text{-}\mathrm{H}\text{-}\mathrm{CZ}$ maps $\ket{\psi}\ket{+}$ to $\ket{+}\ket{\psi}$, transferring the left-site qubit state to the right site and resetting the left site to $\ket{+}$.
    In~\Cref{fig:state_transfer_absent}, both CZ gates act as the identity because the interaction partner is absent, and only the H gate is applied to the right-site qubit, leaving it in the default $\ket{0}$ state.}
  \label{fig:state_transfer_circuit}
\end{figure}

The parallel logical state-transfer operation is a key component of the proposed operation, enabling the parallel and coherent transfer of quantum states efficiently.
The basic idea is to decompose the two CNOT gates used in SWAP-LDU into a sequence of CZ- and H-gate operations.
As shown in \Cref{fig:state_transfer_circuit}, this operation is implemented by applying a parallel CZ operation, a parallel H-gate operation, and another parallel CZ operation in this order.
Here, we consider the two cases of the parallel logical state-transfer operation.
\begin{enumerate}[label=\textbf{Case \arabic*.},leftmargin=*]
  \renewcommand{\labelenumi}{\textbf{Case \arabic{enumi}.}}
  \item The left-site qubit is present in the state $\ket{\psi} = \alpha\ket{0} + \beta\ket{1}$.
        Since the parallel H-gate operation applies the H gate to both the left-site and right-site qubits, the gate sequence maps $\ket{\psi}\ket{+}$ to $\ket{+}\ket{\psi}$.
        Specifically, the first parallel CZ operation maps $\ket{\psi}\ket{+}$ to $\alpha\ket{0}\ket{+} + \beta\ket{1}\ket{-}$, the parallel H-gate operation maps this state to $\alpha\ket{+}\ket{0} + \beta\ket{-}\ket{1}$, and the second parallel CZ operation maps it to $\ket{+}\ket{\psi}$.
        As a result, the state of the left-site qubit is transferred to the right-site qubit, and the left-site qubit is reset to the $\ket{+}$ state.
  \item The left-site qubit is absent.
        In this case, both parallel CZ operations act as the identity on the right-site qubit because its interaction partner is absent~\cite{cite:CZ_PRL}, and the parallel H-gate operation is applied only to the existing right-site qubit, transforming its state from $\ket{+}$ to $\ket{0}$.
        As a result, the right-site qubit remains in the default $\ket{0}$ state.
\end{enumerate}

It should be noted that we implement a one-way state transfer using the gate operations described above, instead of the physical SWAP gate, for the atom reloading protocol.
In typical neutral-atom systems, the SWAP gate is implemented by physically swapping the atoms because it is more efficient than decomposing the SWAP gate into multiple CZ and H-gate operations.
This physical SWAP gate is, however, not suitable for the atom reloading protocol because it swaps the defects between the left sites and right sites of the entanglement array.
This makes the reloading protocol, which reloads the right sites of the entanglement array, meaningless because the defects do not get replenished.
To avoid this issue, we employ the one-way state transfer using the state-transfer operation, which only transfers the quantum states of the atoms without physically swapping them.
This allows the reloading protocol to replenish defects in the storage array with atoms from the entanglement array, achieving the intended effect of the operation.

To account for atom loss errors introduced by the additional gate operations, we introduced $p_\mathrm{op}$ as the error rate of the proposed operation, which is the probability that the operation fails to replenish the defects in the storage array.
This error may cause the existing atoms to be lost during the operation because the operation will replace the existing atoms with new ones if they exist.
We assume that the error occurs independently for each atom in the storage array.

The proposed operation allows us to adaptively choose which rows and columns $I,J$ of the storage array to replenish based on defect information.
In the method of Chiu et al.~\cite{cite:3000_coherent}, the rows and columns to be replenished must be predetermined because the operation discards all the data in the specified rows and columns.
To avoid data loss, the data in the affected atoms must be transferred to other atoms before the operation.
Hence, the predetermined pattern is necessary to ensure that the data in the affected atoms can be transferred to other atoms.
In the proposed operation, the data in the existing atoms is mostly preserved.
This allows us to adaptively choose $I,J$ without considering the data in the existing atoms, which is a significant advantage of the proposed operation.
Even if it is unknown whether the atoms in the specified rows and columns are present or absent, the proposed operation can still be executed without data loss.

The tradeoff of the proposed operation is that it introduces additional operational noise that may affect the existing atoms.
The parallel CZ operation has a reported fidelity of around $99.5\%$~\cite{cite:CZ_parallel}, which is equivalent to an error rate of $0.5\%$.
The total error rate of the proposed operation is around $1.0\%$ because the parallel logical state-transfer operation consists of two parallel CZ operations.
The model partially counts the errors through $p_\mathrm{op}$, although it does not fully capture all the operational noise introduced by the additional gate operations.
While this error rate is not negligible, the errors occur mainly on a limited number of existing atoms selected by $I,J$.
The improved filling rate of the storage array can reduce the number of defects, which in turn reduces errors due to absent atoms.

In the context of quantum error correction, the proposed operation may replace one syndrome-extraction round with two additional CZ gates on the selected atoms.
The improved filling of the storage array can potentially reduce the overall logical error rate by decreasing the number of defects that lead to errors.
It depends on the specific quantum error correction code whether the reduction in error rate outweighs the additional errors introduced by the proposed operation or not.

\subsection{Parameter Optimization of the Proposed Operation}

To determine a preferred $(I,J)$, the proposed defect-aware planner takes a greedy approach to find a solution.
For each time $t$, the proposed defect-aware planner tries to choose a feasible $(I,J)$ that maximizes
\begin{equation}
  \sum^{H_s}_{i = 1} \sum^{W_s}_{j = 1} A^{(t+1)}_{i,j}.
\end{equation}
In other words, we optimize this objective myopically at each time step.

We formulate this objective as follows.
Let $X$ be the state of the storage array after the atom loss detection procedure, i.e., $X = A'^{(t)}$.
Then, the optimization problem can be represented as
\begin{equation}\label{eq:modeling_optimize_problem}
  \begin{aligned}
     & \text{maximize}
     & \sum_{i \in I} \sum_{j \in J} ((1 - p_{\mathrm{op}}) - X_{i,j}),              \\
     & \text{subject to}
     & |I| \leq H_p,                                                   |J| \leq W_p. \\
  \end{aligned}
\end{equation}
In this formulation, we want to find an $(I,J)$ that maximizes the expected number of atoms in the storage array after the operation, such that the sizes of $I$ and $J$ do not exceed those of the entanglement array.

The optimization problem in \Cref{eq:modeling_optimize_problem} is an instance of the dense bipartite subgraph selection problem with two-sided cardinality constraints.
This problem is closely related to the maximum edge biclique problem~\cite{cite:maximum_edge_biclique}.
Our contribution lies in its identification in the context of atom reloading, together with a solver tailored to the real-time planning budget.

The optimization problem in \Cref{eq:modeling_optimize_problem} is NP-hard, which suggests that finding an exact solution efficiently for large instances is computationally challenging.
NP-hardness follows from known results on related problems.
For completeness, we give a direct reduction from the balanced complete bipartite subgraph problem~\cite{cite:garey_johnson}.
For simplicity, we assume that $p_{\mathrm{op}} = 0$.
Let $G = (V_1 \cup V_2, E)$ be a bipartite graph with vertex sets $V_1$ and $V_2$ and edge set $E$.
The balanced complete bipartite subgraph problem is to determine whether there exists a complete bipartite subgraph $(V_1' \cup V_2', E')$ such that $|V_1'| = |V_2'| = k$ for a given integer $k$.
To reduce the balanced complete bipartite subgraph problem to the optimization problem in \Cref{eq:modeling_optimize_problem}, we construct a binary matrix $X$ of size $|V_1| \times |V_2|$ such that $X_{i,j} = 0$ if and only if $(v_i, v_j) \in E$ for $v_i \in V_1$ and $v_j \in V_2$.
Then, we set $H_p = W_p = k$.
We can see that there exists a complete bipartite subgraph $(V_1' \cup V_2', E')$ such that $|V_1'| = |V_2'| = k$ if and only if there exists an $(I,J)$ such that $|I| = |J| = k$ and $\sum_{i \in I} \sum_{j \in J} (1 - X_{i,j}) = k^2$.
Therefore, the optimization problem in \Cref{eq:modeling_optimize_problem} is NP-hard.

There are two approaches to solve the optimization problem in \Cref{eq:modeling_optimize_problem}.
The first is an exact solver that finds the optimal solution, and the second is a heuristic solver that finds a locally optimal solution.
The exact solver can find the optimal solution with a tradeoff of longer execution time that scales exponentially with the problem size.
The heuristic solver can find a locally optimal solution in a reasonable time, but it does not guarantee the optimality of the solution.
We primarily focus on the heuristic solver in this work, while we provide a brief description of the exact solver for comparison.

\subsection{Exact Solver}
The exact solver finds the optimal solution to the optimization problem (\ref{eq:modeling_optimize_problem}) by reformulating it as a mixed integer linear programming (MILP) problem and solving it using an MIP solver~\cite{cite:GLPK,cite:SCIP}.
We use the idea of McCormick relaxation~\cite{cite:mccormick} to linearize the objective function and constraints.
We introduce binary variables $y_i \in \{0,1\}$ and $z_j \in \{0,1\}$ indicating whether row $i$ and column $j$ are selected.
We also introduce the auxiliary variables $w_{i,j}, \mu_b,$ and $\xi_b$ to linearize the optimization problem.
The MILP formulation is given as
\begin{equation}\label{eq:mip_formulation}
  \begin{aligned}
     & \text{maximize}
     & \sum_{i=1}^{H_s} \sum_{j=1}^{W_s} (1 - X_{i,j})\, w_{i,j}
    \;-\; p_{\mathrm{op}} \sum_{b=0}^{W_p} b\, \xi_b,                                                                \\
     & \text{subject to}
     & \sum_{i=1}^{H_s} y_i \leq H_p,                                                                                \\
     &                                                           & \sum_{j=1}^{W_s} z_j \leq W_p,                    \\
     &                                                           & w_{i,j} \leq y_i,                                 \\
     &                                                           & w_{i,j} \leq z_j,                                 \\
     &                                                           & w_{i,j} \geq y_i + z_j - 1,                       \\
     &                                                           & \sum_{b=0}^{W_p} \mu_b = 1,                       \\
     &                                                           & \sum_{j=1}^{W_s} z_j
    = \sum_{b=0}^{W_p} b\, \mu_b,                                                                                    \\
     &                                                           & \sum_{b=0}^{W_p} \xi_b
    = \sum_{i=1}^{H_s} y_i,                                                                                          \\
     &                                                           & \xi_b \leq H_p\, \mu_b,                           \\
     &                                                           & y_i, z_j, \mu_b \in \{0,1\}, w_{i,j} \in [0,1],\;
    \xi_b \geq 0.
  \end{aligned}
\end{equation}
With the constraints, the auxiliary variables $w_{i,j}, \mu_b,$ and $\xi_b$ are uniquely determined by the original binary variables $y_i$ and $z_j$.
The intuitive meaning of these auxiliary variables is as follows.
\begin{itemize}
  \item $w_{i,j}$ represents the conjunction of $y_i$ and $z_j$, indicating whether both row $i$ and column $j$ are selected.
  \item $\mu_b$ represents the number of selected columns using a one-hot representation. $\mu_b = 1$ holds if and only if exactly $b$ columns are selected. Other values are equal to zero.
  \item $\xi_b$ amplifies $\mu_b$ to reflect the number of selected rows. The constraints force $\xi_b$ to satisfy
        \begin{equation}
          \xi_b =  \left( \sum_{i=1}^{H_s} y_i \right) \mu_b
        \end{equation}
        for any $b$.
        This ensures that
        \begin{equation}
          \sum_{b=0}^{W_p} b\, \xi_b = \left( \sum_{i=1}^{H_s} y_i \right) \left( \sum_{j=1}^{W_s} z_j \right)
        \end{equation}
        holds.
\end{itemize}
Since the objective function and all constraints are linearized, this formulation can be solved to optimality using an MIP solver~\cite{cite:GLPK,cite:SCIP}.
While the exact solver may not be practical for large instances due to its performance, this approach allows us to evaluate the performance of the heuristic solver by comparing its solutions to the optimal solutions obtained from the exact solver.

\subsection{Heuristic Solver}

\begin{figure*}[t]
  \centering
  \includegraphics[keepaspectratio, width=0.98\linewidth]{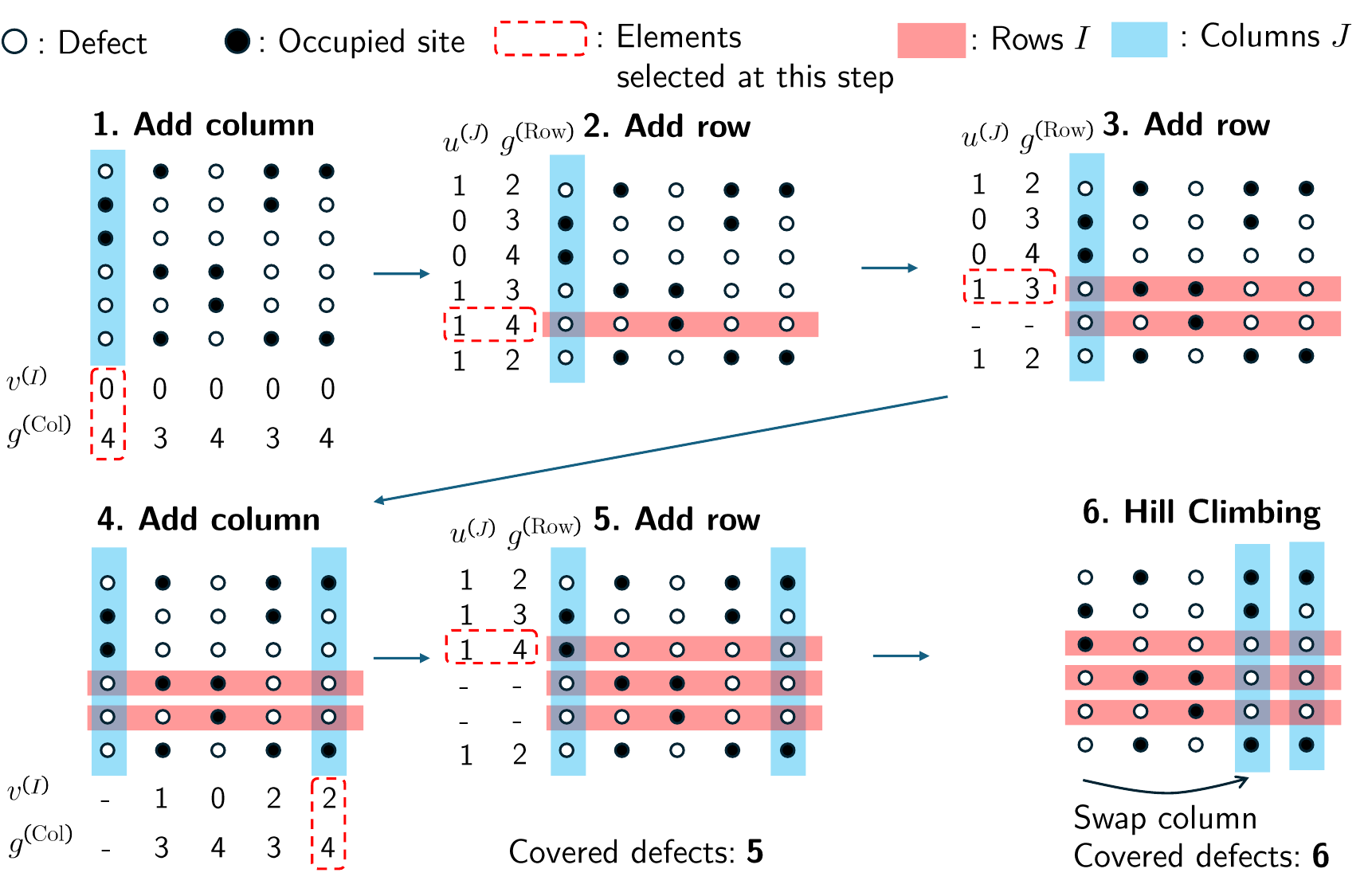}
  \caption{Step-by-step execution of the proposed defect-aware planner on a $6 \times 5$ instance with $H_p = 3$ and $W_p = 2$.
    Steps~1 to~5 show the greedy construction of \Cref{alg:greedy_initial}.
    Whether a row or a column is added is decided by the condition $|I| \cdot W_p < |J| \cdot H_p$, which yields the order column, row, row, column, row for this instance.
    In each step, the elements with the largest $u^{(J)}$ or $v^{(I)}$ are chosen to maximize the number of covered defects.
    In case of a tie, the algorithm selects the row or column with the largest total defect count, given by $(g^{\mathrm{(Row)}}_i)$ or $(g^{\mathrm{(Col)}}_j)$, respectively, to encourage greater defect coverage in subsequent steps.
    These steps yield the greedy solution $I = \{3,4,5\}$, $J = \{1,5\}$, which covers $5$ defects.
    Step~6 shows the hill-climbing refinement of \Cref{alg:hill_climbing}.
    Hill climbing replaces column~1, whose $v^{(I)} = 2$ is the smallest within $J$, by column~4, whose $v^{(I)} = 3$ is the largest outside $J$.
    With this replacement, the refined solution covers $6$ defects.
    The refined solution is globally optimal for this instance because this number of covered defects is equal to the upper bound $H_p W_p = 6$.
    Note that this instance is chosen so that the hill-climbing step can make an improvement.
    As reported in \Cref{sec:convergence_evaluation}, the greedy solution is already nearly optimal in the majority of the instances we evaluated.
    The improvement shown here is therefore larger than the average gain.
  }
  \label{fig:greedy_example}
\end{figure*}

The proposed defect-aware planner heuristically finds a locally optimal solution to the optimization problem (\ref{eq:modeling_optimize_problem}), as shown in \Cref{fig:greedy_example}.
The proposed defect-aware planner consists of two steps: a greedy step that constructs a high-quality solution $s_0$, and a hill-climbing step that certifies the local optimality of $s_0$ and refines it when an improvement is possible.
As shown in \Cref{sec:convergence_evaluation}, the greedy solution is already nearly optimal in most instances.
The hill-climbing step certifies this at negligible cost and refines the solution otherwise.

\subsubsection[Greedy Algorithm]{Greedy Algorithm}

We generate the greedy solution $s_0$ using a greedy algorithm, as shown in \Cref{alg:greedy_initial}.
For a given solution $(I,J)$, we define two vectors $u^{(J)}, v^{(I)}$ whose elements are defined for each row $i$ and column $j$ as
\begin{equation}
  \begin{aligned}
    u^{(J)}_i & = \sum_{j \in J} (1 - X_{i,j}), \\
    v^{(I)}_j & = \sum_{i \in I} (1 - X_{i,j}). \\
  \end{aligned}
\end{equation}
Similarly, we define $g^{\mathrm{(Row)}}_i$ and $g^{\mathrm{(Col)}}_j$ as the number of empty cells in row $i$ and column $j$, respectively, as
\begin{equation}\label{eq:g_def}
  \begin{aligned}
    g^{\mathrm{(Row)}}_i & = \sum_{j=1}^{W_s} (1 - X_{i,j}), \\
    g^{\mathrm{(Col)}}_j & = \sum_{i=1}^{H_s} (1 - X_{i,j}). \\
  \end{aligned}
\end{equation}
The algorithm starts with empty sets $I$ and $J$ and iteratively adds one row or one column at each step.
At each step, the algorithm first determines whether to add a row or a column so that the ratio $|I| / |J|$ approaches the target ratio $H_p / W_p$.
If $|I| \cdot W_p < |J| \cdot H_p$, the algorithm adds a row because the current ratio is smaller than the target ratio.
Otherwise, it adds a column.
After determining the type, the algorithm selects the row or column with the largest number of defects that can be replenished, with ties broken by the largest number of defects in the row or column, i.e., $\arg\max_{i \notin I} u^{(J)}_i \times (W_s+1) + g^{\mathrm{(Row)}}_i$ or $\arg\max_{j \notin J} v^{(I)}_j \times (H_s+1) + g^{\mathrm{(Col)}}_j$.
The algorithm then adds the selected row or column to $I$ or $J$, respectively.
This process is repeated until $|I| = H_p$ and $|J| = W_p$.
After this construction, the algorithm removes selected rows with $u^{(J)}_i = 0$ and selected columns with $v^{(I)}_j = 0$.
Removing them preserves all covered defects while reducing the number of addressed sites because these rows and columns cover no defects within the selected set $I \times J$.
Consequently, this filtering cannot decrease the objective function in \Cref{eq:modeling_optimize_problem} for $p_{\mathrm{op}} \geq 0$.
The two retained sets, denoted $I_+$ and $J_+$, are computed from the same pair $(I,J)$ before filtering and returned as the greedy solution $s_0$.

\begin{algorithm}[t]
  \caption[Greedy Algorithm]{Greedy Algorithm}
  \label{alg:greedy_initial}
  \begin{algorithmic}[1]
    \Require Defect matrix $X$, preparation size $H_p$, $W_p$, storage size $H_s$, $W_s$
    \Ensure Greedy solution $(I_+, J_+)$
    \State $I \gets \emptyset$, $J \gets \emptyset$
    \While{$|I| < H_p$ \textbf{or} $|J| < W_p$}
    \If{$|I| \cdot W_p < |J| \cdot H_p$}
    \State $i^* \gets \arg\max_{i \notin I} u^{(J)}_i \times (W_s+1) + g^{\mathrm{(Row)}}_i$ \Comment{$g^{\mathrm{(Row)}}_i$ is computed from \Cref{eq:g_def}}
    \State $I \gets I \cup \{i^*\}$
    \Else
    \State $j^* \gets \arg\max_{j \notin J} v^{(I)}_j \times (H_s+1) + g^{\mathrm{(Col)}}_j$ \Comment{$g^{\mathrm{(Col)}}_j$ is computed from \Cref{eq:g_def}}
    \State $J \gets J \cup \{j^*\}$
    \EndIf
    \EndWhile
    \State $I_+ \gets \{i \in I \mid u^{(J)}_i > 0\}$
    \State $J_+ \gets \{j \in J \mid v^{(I)}_j > 0\}$
    \State \Return $(I_+, J_+)$
  \end{algorithmic}
\end{algorithm}

\subsubsection{Algorithmic Optimization}

We consider the incremental update of $u^{(J)}$ and $v^{(I)}$ because it is computationally expensive to recalculate them from scratch at each greedy step.
If we calculate $u^{(J)}$ and $v^{(I)}$ from scratch, the time complexity is $\mathrm{O}(H_s W_s)$.
Now, we show that adding one element to $I$ or $J$ can be done in $\mathrm{O}(H_s + W_s)$ time if the vectors $u^{(J)}$ and $v^{(I)}$ are already known.
We consider the case in which one row index, denoted $i^*$, is added to $I$.
In this case, the following equation holds for every column $j$.
\begin{equation}\label{eq:update_relation}
  v^{(I \cup \{i^*\})}_{j} = v^{(I)}_j + (1 - X_{i^*,j}).
\end{equation}
The right-hand side of \Cref{eq:update_relation} can be computed in $\mathrm{O}(1)$ time if the value of $v^{(I)}_j$ is known.
Hence, the value of $v^{(I \cup \{i^*\})}$ can be calculated in $\mathrm{O}(W_s)$ time.
We can take the same approach for the case in which one column index is added to $J$.
Therefore, the new values of the vectors $u^{(J)}, v^{(I)}$ can be computed in $\mathrm{O}(H_s + W_s)$ time.

With this incremental update, we can show that \Cref{alg:greedy_initial} runs in $\mathrm{O}(H_s W_s + (H_p + W_p)(H_s + W_s))$ time.
The precomputation of $g^{\mathrm{(Row)}}, g^{\mathrm{(Col)}}$ takes $\mathrm{O}(H_s W_s)$ time.
Since the algorithm adds one element to $I$ or $J$ at each step, the number of iterations on lines 2--10 is at most $H_p + W_p$.
Each iteration requires $\mathrm{O}(H_s + W_s)$ time to find the best row or column to add, and $\mathrm{O}(H_s + W_s)$ time to update the vectors $u^{(J)}, v^{(I)}$.
Hence, the total time complexity of \Cref{alg:greedy_initial} is $\mathrm{O}(H_s W_s + (H_p + W_p)(H_s + W_s))$.
Since $H_p \leq H_s$ and $W_p \leq W_s$, this time complexity can be simplified to $\mathrm{O}((H_s + W_s)^2)$.
The final filtering takes $\mathrm{O}(H_p + W_p)$ time using the already computed vectors $u^{(J)}, v^{(I)}$.
If these vectors are reused in the subsequent hill-climbing step, they must be updated to reflect the removed rows and columns.
Applying the incremental updates for these removals takes at most $\mathrm{O}(H_p W_s + W_p H_s)$ time, which does not change the overall asymptotic complexity.

\subsubsection{Hill Climbing Refinement}

\begin{algorithm}[t]
  \caption{Hill Climbing}
  \label{alg:hill_climbing}
  \begin{algorithmic}[1]
    \Require Initial solution $s$, evaluation function $E$, neighborhood function $L$
    \Ensure locally optimal solution $s$
    \While{\textbf{true}} \Comment{Repeat until a local optimum is reached}
    \State $s' \gets \arg\max_{s'' \in L(s)} E(s'')$
    \If{$E(s') \leq E(s)$}
    \State \Return $s$ \Comment{Local optimum reached}
    \EndIf
    \State $s \gets s'$
    \EndWhile
  \end{algorithmic}
\end{algorithm}

The proposed defect-aware planner improves the quality of the greedy solution $s_0$, which is used as the initial solution, using hill climbing.
Hill climbing~\cite{cite:local_search} is a local search algorithm that iteratively improves the current solution with respect to an evaluation function $E$, as shown in \Cref{alg:hill_climbing}.
Starting from an initial solution $s$, the algorithm repeatedly moves to the best solution $s^\prime$ in the neighborhood $L(s)$, a set of solutions obtained by slightly modifying the current solution $s$.
The algorithm terminates when no better solution exists in the neighborhood, at which point the current solution is a local optimum.

We define the parameters of hill climbing as follows.
The evaluation function $E$ is set to the objective function of \Cref{eq:modeling_optimize_problem}.
For a feasible solution $s = (I_s,J_s)$, let $\overline{I}_s = \{1,\ldots,H_s\} \setminus I_s$ and $\overline{J}_s = \{1,\ldots,W_s\} \setminus J_s$ denote the sets of unselected rows and columns, respectively.
We define the neighborhood function $L$ as
\begin{equation}\label{eq:hill_climbing_neighborhood}
  \begin{split}
    L(s) = {} &
    \left\{ (I_s \cup \{x\}, J_s)
    \;\middle|\; x \in \overline{I}_s,\; |I_s| < H_p
    \right\}            \\
              & {} \cup
    \left\{ (I_s, J_s \cup \{x\})
    \;\middle|\; x \in \overline{J}_s,\; |J_s| < W_p
    \right\}            \\
              & {} \cup
    \left\{ (I_s \setminus \{y\}, J_s)
    \;\middle|\; y \in I_s
    \right\}            \\
              & {} \cup
    \left\{ (I_s, J_s \setminus \{y\})
    \;\middle|\; y \in J_s
    \right\}            \\
              & {} \cup
    \left\{ ((I_s \setminus \{y\}) \cup \{x\}, J_s)
    \;\middle|\; x \in \overline{I}_s,\; y \in I_s
    \right\}            \\
              & {} \cup
    \left\{ (I_s, (J_s \setminus \{y\}) \cup \{x\})
    \;\middle|\; x \in \overline{J}_s,\; y \in J_s
    \right\}.
  \end{split}
\end{equation}
This neighborhood consists of three types of moves, add, remove, and change, applied to either the row set $I_s$ or the column set $J_s$.
An add move inserts one unselected index subject to the capacity constraint, a remove move deletes one selected index, and a change move replaces one selected index with an unselected index.

Under this definition, the update of the current solution $s$ to the neighboring solution $s^\prime$ can be done in $\mathrm{O}(H_s + W_s)$ time.
Each move adds or removes one element, or replaces one element with another, in either $I_s$ or $J_s$.
Adding one element to $I_s$ or $J_s$ can be done in $\mathrm{O}(H_s + W_s)$ time.
Removing one element from $I_s$ or $J_s$ can also be done in $\mathrm{O}(H_s + W_s)$ time because it updates the values of the vectors $u^{(J)}, v^{(I)}$ similarly to adding one element.
Hence, the update of the current solution $s$ to the neighboring solution $s^\prime$ can be done in $\mathrm{O}(H_s + W_s)$ time.

We now discuss the computation time optimization of the search procedure in the proposed defect-aware planner.
The best neighboring solution can be found in $\mathrm{O}(H_s + W_s)$ time if $u^{(J)}, v^{(I)}$ are given.
We consider the case in which we replace one element of $I$.
Let $\overline{I}$ be the set of row indices that are not in $I$.
Let $i_1 \in I$ be the element that minimizes $u^{(J)}_{i_1}$.
Similarly, let $i_2 \in \overline{I}$ be the element that maximizes $u^{(J)}_{i_2}$.
Then, the best neighboring solution is represented as
\begin{equation}
  I \cup \{i_2\} \setminus \{i_1\}.
\end{equation}
Finding such $i_1, i_2$ takes $\mathrm{O}(H_s)$ time.
We can take a similar approach for the case in which we replace one element of $J$.
Hence, for the replacement neighboring solutions, the best neighboring solution can be found in $\mathrm{O}(H_s + W_s)$ time.
The addition and removal neighboring solutions can be found in $\mathrm{O}(H_s + W_s)$ time using similar approaches.
Therefore, the best neighboring solution can be found in $\mathrm{O}(H_s + W_s)$ time.

The proposed defect-aware planner runs in $\mathrm{O}(H_s W_s + (H_p + W_p + K)(H_s + W_s))$ time with these optimizations, where $K$ is the number of iterations until convergence.
Since each iteration strictly improves the objective and the feasible space is finite, the algorithm terminates at a local optimum.
If a deterministic runtime bound is required, the number of iterations can be capped at the cost of losing the guarantee of local optimality.
In our evaluation, the algorithm always runs until convergence.

\section{Numerical Evaluation}
We evaluated the proposed method by comparing it with the conventional method~\cite{cite:3000_coherent}, or the baseline method, and several other methods with different planning strategies.
\Cref{table:method_comparison} summarizes the operation and the planner used in each method.
\begin{enumerate}
  \item Baseline method~\cite{cite:3000_coherent}: This method replenishes the storage array with atoms in a predetermined pattern as in Chiu et al.~\cite{cite:3000_coherent}, ignoring defect information.
  \item Proposed method: This method replenishes the storage array with atoms using the proposed defect-aware planner based on defect information.
  \item Greedy method: This method uses only the greedy algorithm in \Cref{alg:greedy_initial}.
        This method does not perform the local search refinement and is used to evaluate the effectiveness of the local search in the proposed method.
  \item Simple greedy method: This method selects the rows $I$ and columns $J$ with the largest number of defects in each row $g^{\mathrm{(Row)}}_i$ and column $g^{\mathrm{(Col)}}_j$, independently of each other, i.e., without taking into account the interaction between the selected rows and columns.
        This method is a simpler version of the greedy method and is used to evaluate the effectiveness of the greedy algorithm in the proposed method.
  \item MIP method: This method finds the optimal solution to the optimization problem (\ref{eq:modeling_optimize_problem}) using an MIP solver.
        This method uses the same atom-reloading operation as the proposed method, but it uses an MIP solver to find the optimal solution instead of a locally optimal solution.
        We compared two open-source solvers, namely SCIP~\cite{cite:SCIP} and GLPK~\cite{cite:GLPK}, for the MIP method.
        For the MIP method, we terminated the evaluation for a given array size and noise condition once the cumulative planning time exceeded the total planned number of calls ($100$ trials $\times$ $100$ operations) times $100$ ms, corresponding to an average planning-time budget of $100$ ms per call.
        This cumulative budget of $1{,}000$ seconds allows individual planning calls to exceed $100$ ms.
        Incomplete settings and larger sizes for that noise condition are excluded from the reported results.
  \item Row-by-row method~\cite{cite:cycle_time_neutral_atom1, cite:neutral_atom_coherent_reuse}: This method replenishes the storage array with atoms row by row using a fixed row order and a defect-aware column selection shown in \Cref{eq:row_by_row_indices}.
        This method is included to evaluate the performance of a simple, defect-aware strategy.
\end{enumerate}

All methods are simulated under the same state-update rule in \Cref{eq:state_update}.
For a given $(I,J)$, the discard-based operation of Chiu et al.~\cite{cite:3000_coherent} and the proposed coherent operation yield the same array state.
Consequently, the filling-rate differences reported below are attributable solely to the planning policy that selects $(I,J)$.
The role of the proposed coherent operation is to make defect-aware selection physically admissible without data loss.
Its cost, the additional gate noise, is assessed separately in \Cref{sec:coherent_operation} and does not appear in the filling-rate metric.

\begin{table}[t]
  \centering
  \caption{Comparison of the evaluated methods.}\label{table:method_comparison}
  \setlength{\tabcolsep}{3pt}
  \begin{tabular}{llll}
    \toprule
    Method                                                                            & Operation     & Planner                                            & Defect-aware \\
    \midrule
    Baseline~\cite{cite:3000_coherent}
                                                                                      & Discard-based & Fixed pattern                                      & No           \\
    Proposed                                                                          & Coherent      & Greedy (\Cref{alg:greedy_initial}) + Hill climbing & Yes          \\
    Greedy                                                                            & Coherent      & Greedy (\Cref{alg:greedy_initial})                 & Yes          \\
    Simple greedy                                                                     & Coherent      & Greedy (independent)                               & Yes          \\
    MIP                                                                               & Coherent      & Exact                                              & Yes          \\
    Row-by-row~\cite{cite:cycle_time_neutral_atom1, cite:neutral_atom_coherent_reuse} & Coherent      & Fixed pattern                                      & Yes          \\
    \bottomrule
  \end{tabular}
\end{table}

For simplicity, we assume that one reloading procedure takes $1$ unit time regardless of the number of atoms reloaded simultaneously.
Note that the row-by-row atom reloading method is included only in the filling-rate table (\Cref{table:filling_rate_n3}) because the row-by-row method showed a lower filling rate than the other methods under the evaluated settings.
Our numerical evaluation was performed on a machine with the specifications listed in \Cref{table:experimental_environment}.
The reported results are averages over $100$ trials for each completed setting, and the error bars in the figures denote the sample standard deviation across trials.
All methods share the same random seed for a given setting.

\begin{table}[t]
  \centering
  \caption{Experimental environment.}\label{table:experimental_environment}
  \begin{tabular}{ll}
    \toprule
    CPU      & Intel Core\texttrademark{} i5-12400F \\
    Memory   & 24GB                                 \\
    OS       & Ubuntu 24.04.3 LTS                   \\
    Compiler & g++ 13.3.0 (\texttt{-std=c++17 -O2}) \\
    SCIP     & v.10.0.3 (LP solver: SoPlex 8.0.3)   \\
    GLPK     & v.5.0-1build2                        \\
    \bottomrule
  \end{tabular}
\end{table}

We evaluated the methods in terms of the following metrics.
\begin{enumerate}
  \item The filling rate.
  \item The execution time of the planner.
  \item The number of iterations until convergence.
  \item The transition of the filling rate over time.
\end{enumerate}

\begin{table}[t]
  \centering
  \caption{Parameter settings.}\label{table:parameter_settings}
  \begin{tabular}{ll}
    \toprule
    System Size Scale Factor & $1 \leq n \leq 15$   \\
    Storage Array Size                       & $H_s = 12n, \quad W_s = 30n$         \\
    Entanglement Array Size                  & $H_p = H_s / 3, \quad W_p = W_s / 2$ \\
    Total Time Steps                         & $T = 100$                            \\
    Burn-in Steps                            & $C = 20$                             \\
    Initial State of Storage Array           & $A^{(0)}$ is fully occupied          \\
    Error Rates                              & $p = 0.004, 0.020$                   \\
    \bottomrule
  \end{tabular}
\end{table}

We set the parameters based on the experimental setup of Chiu et al.~\cite{cite:3000_coherent}.
Let $n$ be a positive integer specifying the size scale factor of the system, varying from $1$ to $15$.
Then, we evaluated the methods using the parameters listed in \Cref{table:parameter_settings}.
For simplicity, we assume $p = p_\mathrm{idle} = p_\mathrm{op}$ although the proposed method can be applied to the case where $p_\mathrm{idle} \neq p_\mathrm{op}$.

The low-noise case $p = 0.004$ is calibrated so that the simulated filling rate of the fixed-pattern method ($98.61\%$ at $n = 3$, whose $36 \times 90$ array is comparable in size to the $3{,}000$-qubit system) is slightly higher than the filling rate of $98.5\%$ reported by Chiu et al.~\cite{cite:3000_coherent}.
Under an exponential loss model with the $80$\,ms reload cycle, this loss probability corresponds to an effective single-atom lifetime of approximately $20$\,s.
The high-noise case $p = 0.020$, five times the calibrated value, is a stress-test setting that evaluates the robustness of the methods under loss rates substantially higher than in current experiments.
Under the same exponential loss model, it corresponds to an effective lifetime of approximately $4$\,s.
Such a rate could arise, for example, if gate- and measurement-induced losses during active error correction dominate the background loss.

\begin{figure*}[t]
  \centering
  \begin{subfigure}[t]{0.45\linewidth}
    \includegraphics[keepaspectratio, width=\linewidth]{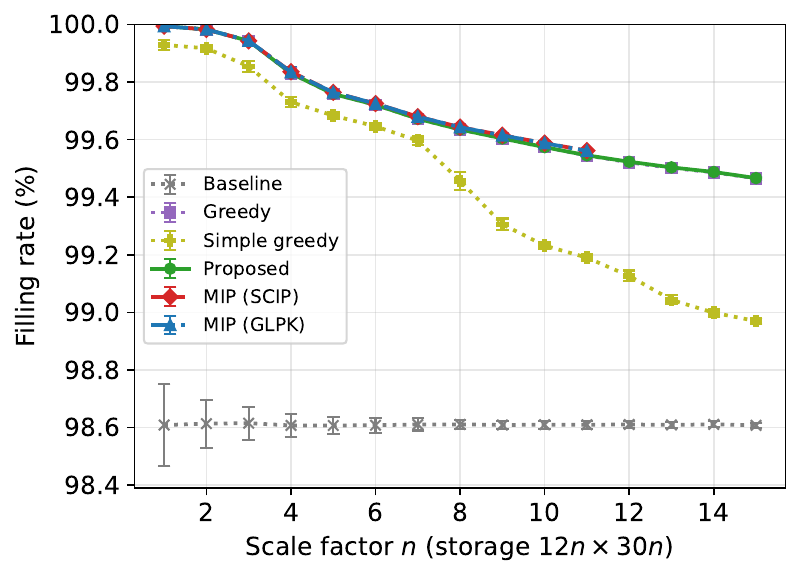}
    \caption{}
    \label{fig:filling_rate_low_noise}
  \end{subfigure}\hfill
  \begin{subfigure}[t]{0.45\linewidth}
    \includegraphics[keepaspectratio, width=\linewidth]{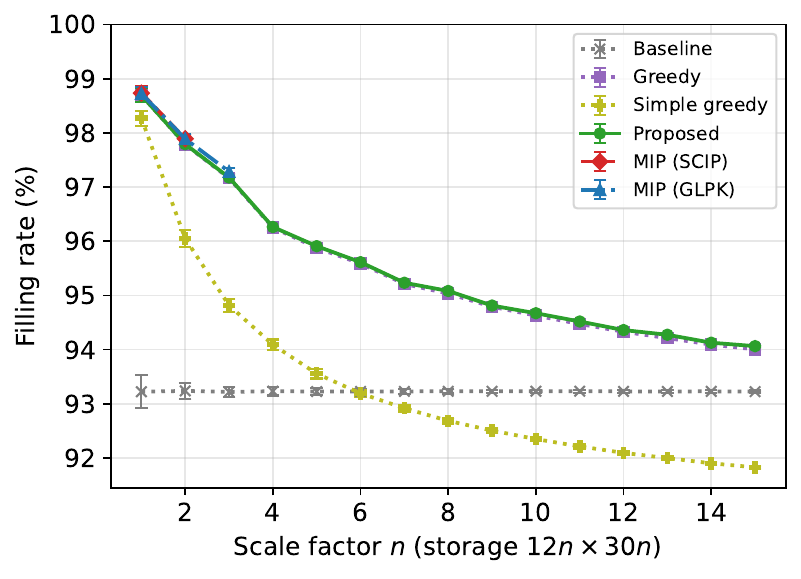}
    \caption{}
    \label{fig:filling_rate_high_noise}
  \end{subfigure}
  \caption{Average filling rate $F$ of the storage array.
    \Cref{fig:filling_rate_low_noise,fig:filling_rate_high_noise} show the results for $p = 0.004$ and $p = 0.020$, respectively.
    The proposed method outperforms the baseline method in both low-noise and high-noise environments.
    The MIP curves in \Cref{fig:filling_rate_low_noise} are shown up to $n = 11$ for both SCIP and GLPK, and those in \Cref{fig:filling_rate_high_noise} up to $n = 2$ for SCIP and $n = 3$ for GLPK, because they failed to complete executions within the time limit of $1{,}000$ seconds.
    Error bars denote the sample standard deviation over $100$ trials and are smaller than the markers in most cases.
    Note that the filling-rate metric does not include the additional gate noise introduced by the proposed coherent operation.
  }
  \label{fig:filling_rate_evaluation}
\end{figure*}

\subsection{Filling Rate} \label{sec:filling_rate_evaluation}
The proposed method improves the filling rate $F$ compared with the baseline method, as shown in \Cref{fig:filling_rate_evaluation}.
\Cref{table:filling_rate_n3} lists the filling rates at $n = 3$, whose $36 \times 90$ storage array is comparable in size to the $3{,}000$-qubit system of Chiu et al.~\cite{cite:3000_coherent}.
For the $p = 0.004$ case shown in \Cref{fig:filling_rate_low_noise}, the proposed method achieved a filling rate of $99.943\%$, while the baseline method achieved $98.615\%$, an improvement of $1.33$ percentage points.
For the $p = 0.020$ case shown in \Cref{fig:filling_rate_high_noise}, the proposed method achieved $97.177\%$, while the baseline method achieved $93.218\%$, an improvement of $3.96$ percentage points.
The baseline filling rate is essentially independent of $n$, because its fixed pattern refreshes every site once per six time steps regardless of the array size.

\begin{table}[t]
  \centering
  \caption{Filling rate $F$ and planning cost at $n = 3$ (storage $36 \times 90$, entanglement $12 \times 90$), averaged over $100$ trials. The $\pm$ values are sample standard deviations across trials. ``Time/step'' is the mean planning time per reload step and ``Worst plan'' is the longest single planner invocation observed. Both measure planning calls only, exclude simulator overhead, and include the burn-in steps. A dash indicates that no complete $100$-trial result is available.}\label{table:filling_rate_n3}
  \setlength{\tabcolsep}{4pt}
  \begin{tabular}{lccc}
    \toprule
    Method                             & $F$ (\%)             & Time/step (ms) & Worst plan (ms) \\
    \midrule
    \multicolumn{4}{l}{\textit{Low-noise case ($p = 0.004$)}}                                    \\
    Baseline~\cite{cite:3000_coherent} & $98.6149 \pm 0.0569$ & $0.002$        & $0.007$         \\
    Proposed                           & $99.9430 \pm 0.0123$ & $0.014$        & $0.043$         \\
    Greedy                             & $99.9430 \pm 0.0123$ & $0.011$        & $0.027$         \\
    Simple greedy                      & $99.8545 \pm 0.0197$ & $0.006$        & $0.014$         \\
    MIP (GLPK)                         & $99.9436 \pm 0.0116$ & $1.681$        & $7.056$         \\
    MIP (SCIP)                         & $99.9430 \pm 0.0121$ & $6.710$        & $492.662$       \\
    Row-by-row                         & $93.3362 \pm 0.2461$ & $0.0005$       & $0.0063$        \\
    \midrule
    \multicolumn{4}{l}{\textit{High-noise case ($p = 0.020$)}}                                   \\
    Baseline~\cite{cite:3000_coherent} & $93.218 \pm 0.099$   & $0.002$        & $0.014$         \\
    Proposed                           & $97.177 \pm 0.081$   & $0.017$        & $0.038$         \\
    Greedy                             & $97.163 \pm 0.078$   & $0.014$        & $0.022$         \\
    Simple greedy                      & $94.816 \pm 0.125$   & $0.007$        & $0.019$         \\
    MIP (GLPK)                         & $97.286 \pm 0.070$   & $12.998$       & $208.265$       \\
    MIP (SCIP)                         & ---                  & ---            & ---             \\
    Row-by-row                         & $69.684 \pm 0.578$   & $0.0016$       & $0.0144$        \\
    \bottomrule
  \end{tabular}
\end{table}

These results show that defect-aware planning, which is enabled by the proposed operation, is more effective in improving the filling rate than the baseline method that ignores defect information.
The baseline method reloads atoms in a predetermined pattern, which may not align with the actual defects in the storage array.
The proposed approach, which searches for a solution to the optimization problem (\ref{eq:modeling_optimize_problem}) based on defect information, can efficiently fill the defects in the storage array.

The simple greedy method isolates how much of this gain requires modeling the interaction between the selected rows and columns.
Selecting rows and columns independently incurs a small loss while the defects are sparse.
For $p = 0.004$, the simple greedy algorithm already falls below the greedy algorithm at $n = 1$, with $99.929\%$ versus $99.994\%$.
Once the defects become dense, however, independent selection degrades sharply.
For $p = 0.020$, the simple greedy algorithm already loses $0.42$ percentage points at $n = 1$, and from $n = 6$ onward it falls below the baseline, reaching $91.828\%$ at $n = 15$ against the baseline's $93.224\%$.
Therefore, the planner that ignores the product structure of $(I,J)$ can be worse than not using defect information at all, meaning that defect awareness alone is not sufficient.

The comparison of the proposed method and the greedy method shows the contribution of the local-search refinement.
In the low-noise case at $n = 3$, the two methods are identical on every trial because the greedy solution is already locally optimal in all instances, as we show later in \Cref{sec:convergence_evaluation}.
The effect of the local-search refinement becomes evident as the instances get harder.
At $n = 15$ the proposed method gains $0.0010$ percentage points for $p = 0.004$ and $0.060$ percentage points for $p = 0.020$.
Thus, the observed difference in mean filling rate is small in absolute terms.
The greedy construction accounts for almost all of the improvement over the baseline.

The comparison of the proposed method and the MIP method indicates that the proposed method achieves average filling rates close to those obtained by the MIP method.
At $n = 3$ the difference in filling rate is $0.0006$ percentage points in the low-noise case and $0.109$ percentage points in the high-noise case with GLPK.
Over the completed MIP settings, the largest differences are $0.0163$ and $0.109$ percentage points in the low-noise and high-noise cases, respectively.
The locally optimal solutions found by the proposed defect-aware planner recover more than $97\%$ of the filling-rate gain achieved by MIP in both cases.

We attribute the drop in filling rate as $n$ increases to the product structure of the operation parameter $(I,J)$.
A defect can be replenished only if its row is in $I$ and its column is in $J$ simultaneously.
As $n$ increases, the defects spread over more distinct rows and columns.
Once the number of such rows or columns exceeds $H_p$ or $W_p$, no single operation can cover all defects.
This coverage limit is reached earlier in the $p=0.020$ case, which explains the larger drop compared with the $p=0.004$ case.

\begin{figure*}[t]
  \centering
  \begin{subfigure}[t]{0.45\linewidth}
    \includegraphics[keepaspectratio, width=\linewidth]{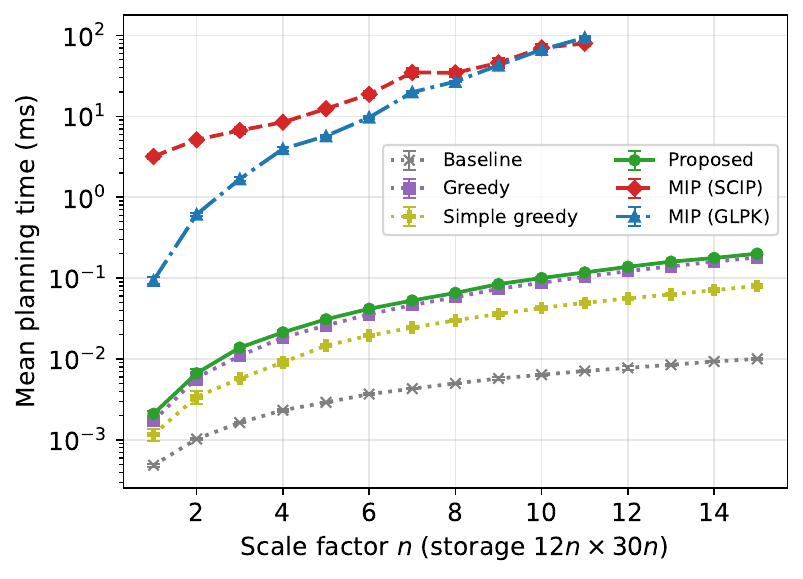}
    \caption{}
    \label{fig:runtime_low_noise}
  \end{subfigure}\hfill
  \begin{subfigure}[t]{0.45\linewidth}
    \includegraphics[keepaspectratio, width=\linewidth]{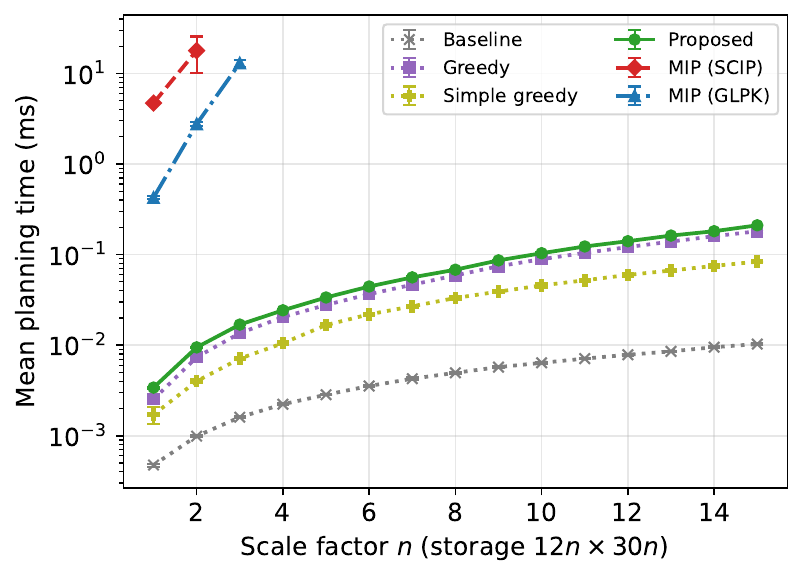}
    \caption{}
    \label{fig:runtime_high_noise}
  \end{subfigure}
  \caption{Average planning time per reload step.
    \Cref{fig:runtime_low_noise,fig:runtime_high_noise} show the results for $p = 0.004$ and $p = 0.020$, respectively.
    The proposed method outperforms the MIP method in both low-noise and high-noise environments.
    The reported time measures planner wall-time for all $100$ steps per trial including burn-in.
    Error bars show the sample standard deviation of the per-trial mean planning times.
    The MIP curves in \Cref{fig:runtime_low_noise} are shown up to $n = 11$ for both SCIP and GLPK, and those in \Cref{fig:runtime_high_noise} up to $n = 2$ for SCIP and $n = 3$ for GLPK.}
  \label{fig:runtime_evaluation}
\end{figure*}

\begin{figure*}[t]
  \centering
  \begin{subfigure}[t]{0.45\linewidth}
    \includegraphics[keepaspectratio, width=\linewidth]{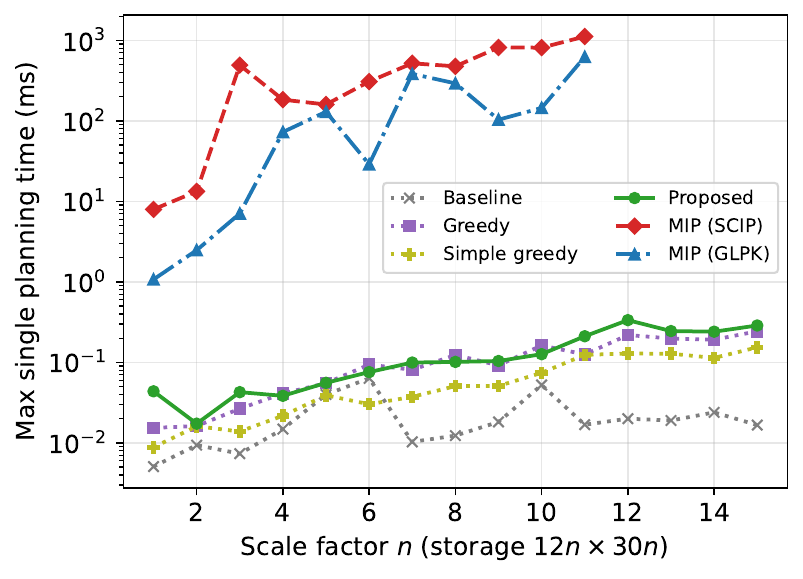}
    \caption{}
    \label{fig:runtime_worst_low_noise}
  \end{subfigure}\hfill
  \begin{subfigure}[t]{0.45\linewidth}
    \includegraphics[keepaspectratio, width=\linewidth]{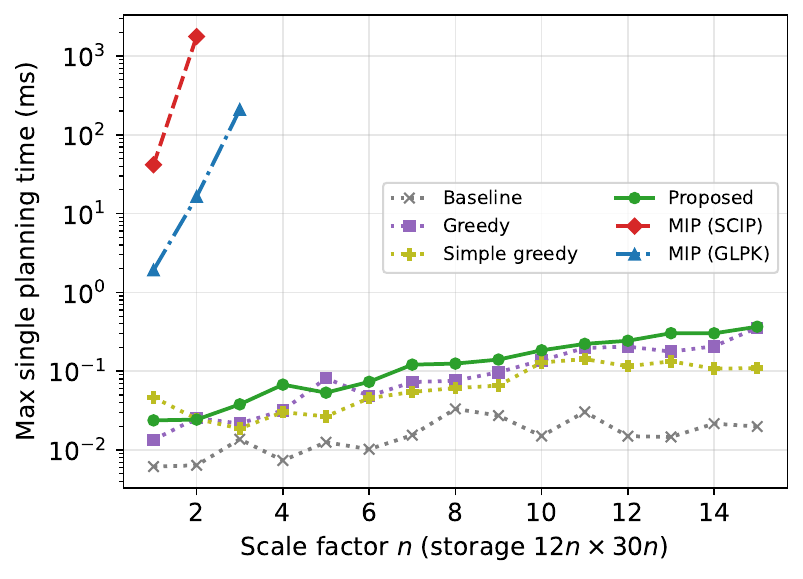}
    \caption{}
    \label{fig:runtime_worst_high_noise}
  \end{subfigure}
  \caption{Maximum planning time per reload operation.
    \Cref{fig:runtime_worst_low_noise,fig:runtime_worst_high_noise} show the results for $p = 0.004$ and $p = 0.020$, respectively.
    The proposed method shows stable behavior, while the MIP method exhibits erratic behavior with occasional long runtimes that exceed the $40$ ms planning budget.
    The MIP curves in \Cref{fig:runtime_worst_low_noise} are shown up to $n = 11$ for both SCIP and GLPK, and those in \Cref{fig:runtime_worst_high_noise} up to $n = 2$ for SCIP and $n = 3$ for GLPK.}
  \label{fig:runtime_worst_evaluation}
\end{figure*}

\subsection[Runtime of the Planner]{Runtime of the Planner}
The proposed defect-aware planner runs faster than the MIP method, as shown in \Cref{fig:runtime_evaluation}.
For the $n = 3$ and $p = 0.004$ case, shown in \Cref{fig:runtime_low_noise}, the proposed method takes $0.014$ ms per reload step, while the greedy method takes $0.011$ ms, and the MIP method takes $6.71$ ms with SCIP and $1.68$ ms with GLPK.
For the $n = 3$ and $p = 0.020$ case, shown in \Cref{fig:runtime_high_noise}, the proposed method takes $0.017$ ms per reload step, while the greedy method takes $0.014$ ms, and the MIP method takes $13.00$ ms with GLPK.
The MIP method with SCIP failed to complete executions within the time limit of $1{,}000$ seconds.

The observed planning times of the proposed method remain well below the control budget, as shown in \Cref{fig:runtime_worst_evaluation}.
For a real-time workload, the worst-case runtime is critical since a single overrun can miss a reload opportunity.
Hence, we also recorded the longest runtime of a single planner invocation over all $100$ trials and all $100$ reload steps.
For the proposed method, the maximum observed planning time is $0.043$ and $0.038$ ms at $n = 3$ under low and high noise, respectively, and remains below $0.37$ ms over the entire range $1 \leq n \leq 15$ in either noise regime.
These results leave more than a factor of $100$ of margin against the $40$ ms planning budget.
On the other hand, the MIP method shows erratic behavior.
At $n = 2$ under high noise, a single SCIP invocation took $1{,}765$ ms, even though its mean planning time of $17.90$ ms per reload step is nominally within budget.
The cumulative planning time reached a budget of $1{,}000$ seconds at $n = 12$ for both SCIP and GLPK under low noise, and at $n = 3$ for SCIP and $n = 4$ for GLPK under high noise.

These results show that the proposed defect-aware planner can find a locally optimal solution within the real-time budget, while the MIP method took a long time, exceeding the real-time budget.
Compared with the greedy method, the local search in the proposed method adds only about $0.003$ ms per reload step at $n = 3$, a negligible overhead for the local-optimality guarantee relative to the $40$ ms planning budget.
Compared with the baseline method, which ignores defect information, the proposed method requires additional computation time to find a locally optimal solution based on defect information.
The absolute overhead is nevertheless small enough to be negligible in practice.
At $n = 15$, the largest size we evaluated, the proposed method takes $0.201$ and $0.211$ ms per reload step under low and high noise, against approximately $0.010$ ms for the baseline in both cases.
The times plotted in \Cref{fig:runtime_evaluation} therefore reflect the cost of selecting the reloading parameters, including greedy construction and hill-climbing refinement for the proposed method.
The maximum planning time of the MIP method exceeds the $40$ ms budget already at $n = 1$ under high noise with SCIP, while the proposed method's mean planning time stays below $0.212$ ms throughout.
The runtime trend therefore suggests that the MIP method is not suitable for larger storage arrays, while the proposed defect-aware planner scales efficiently.

\begin{figure}[t]
  \centering
  \includegraphics[keepaspectratio, width=0.5\linewidth]{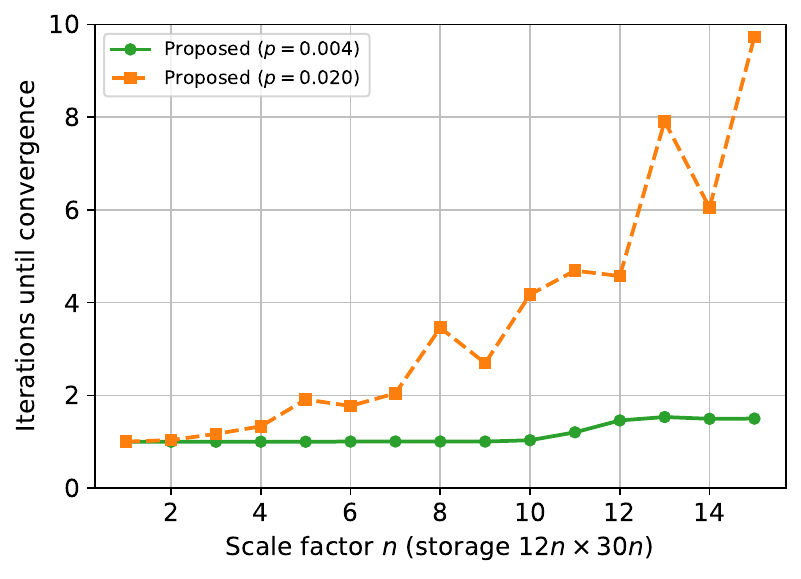}
  \caption{The average number of iterations until convergence for the proposed method.
    An iteration count of $1$ means that the greedy solution is already locally optimal.
    The average iteration count stays below $9.8$ for all $n$.
    The count includes the final neighborhood evaluation that confirms local optimality.
    The average refinement effort increases with the array size and noise level.}
  \label{fig:convergence_evaluation}
\end{figure}

\subsection[Convergence of the Defect-Aware Planner]{Convergence of the Defect-Aware Planner}\label{sec:convergence_evaluation}
The greedy initialization requires little refinement for small arrays, as shown in \Cref{fig:convergence_evaluation}.
For $p = 0.004$ and $n = 3$, the proposed method converges in $1.000$ iterations on average.
For $p = 0.020$ and $n = 3$, the proposed method converges in $1.169$ iterations on average.
Note that we consider the final evaluation process, which confirms local optimality, as one iteration.
Hence, an iteration count of $1$ means that the greedy solution is already locally optimal, and an iteration count greater than $1$ indicates that the local search has improved the greedy solution.
At $n = 3$, these averages indicate little refinement under either noise condition.
The aggregate iteration counts do not report the fraction of locally optimal greedy solutions for each setting.

The number of iterations grows with both the array size and the noise level, which is consistent with the coverage argument given in \Cref{sec:filling_rate_evaluation}.
In the low-noise case the average rises only to $1.499$ at $n = 15$, with a maximum of $1.532$ at $n = 13$.
In the high-noise case, where a single operation can no longer cover all defects, the average grows to its maximum of $9.735$ at $n = 15$.
These results indicate that the proposed defect-aware planner can run in a reasonable time for the considered array sizes and noise levels.

\subsection{Transition of the Filling Rate}
\begin{figure}[t]
  \centering
  \includegraphics[keepaspectratio,width=0.5\linewidth]
  {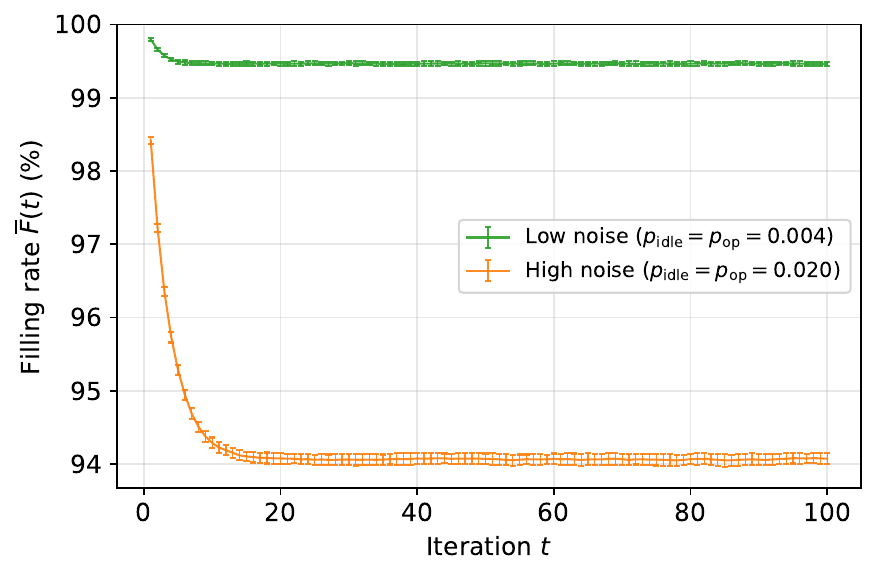}
  \caption{
    Transition of the filling rate $\overline{F}(t)$ for the proposed method at $n=15$, under low noise $p=0.004$ and high noise $p=0.020$.
    Curves and error bars denote the mean and sample standard deviation over $100$ trials, respectively.
    The filling rate becomes stable after the initial burn-in period of $20$ steps.
  }
  \label{fig:filling_rate_time}
\end{figure}

\Cref{fig:filling_rate_time} shows that the filling rate quickly stabilizes after the initial burn-in period of $20$ steps.
Let $\overline{F}(t)$ denote the average filling rate over multiple trials at time step $t$.
Starting from the fully occupied state, $\overline{F}(0)=1$, the mean filling rate decreases during the initial burn-in period.
For $p=0.004$, it reaches $99.467\%$ at $t=10$ and stabilizes thereafter.
For $p=0.020$, it reaches $94.287\%$ at $t=10$ and $94.075\%$ at $t=20$ and stabilizes thereafter.

These results confirm that the burn-in period of $C = 20$ steps is sufficient for the filling rate to stabilize.
If we choose a burn-in period shorter than $C = 20$ steps, the filling rate may not have fully stabilized.
Instability in the filling rate potentially leads to inaccurate assessments of the system's performance.
Hence, in our evaluations and analyses, we adopt a burn-in period of $C = 20$ steps to ensure that the filling rate has stabilized before making any assessments of the system's performance.

\subsection{Limitations}
Our physical model is based on the following assumptions.
\begin{itemize}
  \item The atom loss is detected instantaneously and accurately.
        We note that the inaccurate detection of atom loss does not affect the operation itself but can lead to a lower filling rate because of planning based on inaccurate defect information.
  \item The atom-reloading operation is performed in a single time step regardless of the number of atoms reloaded.
        This abstraction favors massively parallel schemes.
  \item The operational noise introduced by the additional CZ gates is not reflected in the filling-rate metric.
        While we believe that the improvement in the filling rate outweighs the additional operational noise, it has not been verified numerically for its impact on the overall system performance.
  \item The spacing between the atoms in both the storage and entanglement arrays is larger than the minimum tweezer spacing required by AODs.
        This ensures that the tweezers generated by AODs do not interfere with each other.
\end{itemize}

Our defect-aware planner and modeling framework, such as \Cref{eq:modeling_optimize_problem}, have limited modeling capability and do not fully model the effect of the atom-reloading operation on the existing atoms when selecting $(I,J)$.
The assumption of negligible impact on existing atoms simplifies the modeling but may not hold in more realistic scenarios where the operations are executed in shorter time intervals.
Specifically, if multiple atom-reloading operations are performed in relatively shorter time intervals, the effect of the operation on the existing atoms may accumulate and become non-negligible.
The modeling framework can be extended to account for this effect, for instance, by simulating the impact of the atom-reloading operation on the existing atoms within quantum error correction frameworks.
This extension, however, would require a more complex optimization problem and may increase the computational complexity of the overall procedure, which could affect its runtime performance.
Resolving these challenges is left for future work.

\section{Conclusion}

In this work, we proposed a coherent parallel atom-reloading operation that makes defect-aware reloading physically admissible, together with a defect-aware planner that determines its parameters within the real-time control budget.
In numerical experiments on a $36 \times 90$ storage atom array, the average atom filling rate improved from $98.61\%$ to $99.94\%$ under low-noise conditions, and from $93.22\%$ to $97.18\%$ under high-noise conditions.
The proposed defect-aware planner recovers more than $97\%$ of the filling-rate gain achieved by an exact MIP solver over the completed MIP settings, while its maximum observed planning time stays below $0.37$ ms for storage arrays of up to $180 \times 450$ sites, against a $40$ ms budget.
The ablation against the planner that selects rows and columns independently shows that accounting for the product structure of the operation parameters is essential.
The naive planner with independent row and column selection can yield a filling rate below that of the defect-agnostic baseline.
We believe that validating the proposed method on physical hardware and addressing the associated challenges will be important steps for future research.

\section*{Acknowledgements}
The authors would like to thank Dr. Takafumi Tomita of the Institute for Molecular Science for valuable discussions on the physical aspects of this work.

\section*{Author Contributions}
K.A. conceived the study, designed the methodology, developed the software, performed the numerical experiments, and wrote the main manuscript text.
F.I. assisted in refining the manuscript and supervised the project.
All authors reviewed the manuscript.

\section*{Funding}
This work was supported by JSPS KAKENHI Grant Numbers 23K18464, 23K28061, and 26K23814.
K.A. received travel support from ALGO ARTIS Corporation.
ALGO ARTIS Corporation had no role in study design, data collection, analysis, interpretation of results, preparation of the manuscript, or the decision to submit the article for publication.

\section*{Data Availability}
The source code of the proposed method, scripts used to run the numerical evaluation, and the data generated during the experiments are available at \url{https://github.com/kotamanegi/defect-aware-parallel-atom-reloading}.

\section*{Declarations}

\subsection*{Competing interests}
K.A. is a part-time employee of ALGO ARTIS Corporation, which develops products related to combinatorial optimization.
The views expressed in this article are solely those of the authors and do not necessarily reflect the views or positions of ALGO ARTIS Corporation.
The other authors have no competing interests to declare.

\subsection*{Declaration of generative AI and AI-assisted technologies in the manuscript preparation process}
During the preparation of this manuscript, the authors used ChatGPT (OpenAI), Claude (Anthropic), Gemini (Google), Qwen (Alibaba), DeepSeek, and Kimi (Moonshot AI) in order to improve the clarity and readability of the text.
The authors also used these tools to develop the software used in the evaluation and confirm consistency in the results.
After using these tools/services, the authors reviewed and edited the content as needed and take full responsibility for the content of the published article.

% Bibliography generated from citation.bib for submission.

\end{document}